\pdfoutput=1
\documentclass[12pt,a4paper]{article}

\usepackage{subfigure}
\usepackage{makecell}

\usepackage{ifthen} 
\newboolean{pdflatex}
\setboolean{pdflatex}{true} 

\newboolean{articletitles}
\setboolean{articletitles}{true} 

\newboolean{uprightparticles}
\setboolean{uprightparticles}{false} 

\def\paperauthors{LHCb collaboration} 
\def\paperasciititle{Observation of Xib0 -> Xic+ Ds- and Xib- -> Xic0 Ds- decays} 
\def\papertitle{Observation of $\Xibz\to\Xicp\Dsm$ and $\Xibm\to\Xicz\Dsm$ decays} 
\def\paperkeywords{{High Energy Physics}, {LHCb}} 
\def\papercopyright{\the\year\ CERN for the benefit of the LHCb collaboration} 
\def\paperlicence{CC BY 4.0 licence}
\def\paperlicenceurl{https://creativecommons.org/licenses/by/4.0/}

\usepackage[top=1in, bottom=1.25in, left=1in, right=1in]{geometry}

\usepackage{microtype}
\usepackage{lineno}  
\usepackage{xspace} 
\usepackage{caption} 

\usepackage{graphicx}  
\usepackage{color}
\usepackage{colortbl}
\graphicspath{{./figs/}} 

\usepackage{amsmath} 
\usepackage{amssymb}
\usepackage{amsfonts}
\usepackage{upgreek} 

\newcommand*\patchAmsMathEnvironmentForLineno[1]{%
\expandafter\let\csname old#1\expandafter\endcsname\csname #1\endcsname
\expandafter\let\csname oldend#1\expandafter\endcsname\csname
end#1\endcsname
 \renewenvironment{#1}%
   {\linenomath\csname old#1\endcsname}%
   {\csname oldend#1\endcsname\endlinenomath}%
}
\newcommand*\patchBothAmsMathEnvironmentsForLineno[1]{%
  \patchAmsMathEnvironmentForLineno{#1}%
  \patchAmsMathEnvironmentForLineno{#1*}%
}
\AtBeginDocument{%
\patchBothAmsMathEnvironmentsForLineno{equation}%
\patchBothAmsMathEnvironmentsForLineno{align}%
\patchBothAmsMathEnvironmentsForLineno{flalign}%
\patchBothAmsMathEnvironmentsForLineno{alignat}%
\patchBothAmsMathEnvironmentsForLineno{gather}%
\patchBothAmsMathEnvironmentsForLineno{multline}%
\patchBothAmsMathEnvironmentsForLineno{eqnarray}%
}

\usepackage{hyperxmp}

\usepackage[pdftex,
            pdfauthor={\paperauthors},
            pdftitle={\paperasciititle},
            pdfkeywords={\paperkeywords},
            pdfcopyright={Copyright (C) \papercopyright},
            pdflicenseurl={\paperlicenceurl}]{hyperref}

\usepackage[colorinlistoftodos,textsize=scriptsize]{todonotes}

\usepackage[bottom,flushmargin,hang,multiple]{footmisc}

\usepackage[all]{hypcap} 

\usepackage{xspace} 
\usepackage{upgreek}

\def\lhcb   {\mbox{LHCb}\xspace}

\def\MagUp {\mbox{\em Mag\kern -0.05em Up}\xspace}

\ifthenelse{\boolean{uprightparticles}}%
{

 \def\Pmu         {\ensuremath{\upmu}\xspace}

 \def\Ppi         {\ensuremath{\uppi}\xspace}

 \def\Ppsi        {\ensuremath{\uppsi}\xspace}

 \def\PDelta      {\ensuremath{\Delta}\xspace}                 
 \def\PXi         {\ensuremath{\Xi}\xspace}                 
 \def\PLambda     {\ensuremath{\Lambda}\xspace}                 
 \def\PSigma      {\ensuremath{\Sigma}\xspace}                 
 \def\POmega      {\ensuremath{\Omega}\xspace}                 
 \def\PUpsilon    {\ensuremath{\Upsilon}\xspace}
 \let\oldPi\Pi
 \def\PPi         {\ensuremath{\oldPi}\xspace}

 \def\PB      {\ensuremath{\mathrm{B}}\xspace}                 
                  
 \def\PD      {\ensuremath{\mathrm{D}}\xspace}

 \def\PJ      {\ensuremath{\mathrm{J}}\xspace}                 
 \def\PK      {\ensuremath{\mathrm{K}}\xspace}

 \def\Pb      {\ensuremath{\mathrm{b}}\xspace}                 
 \def\Pc      {\ensuremath{\mathrm{c}}\xspace}

 \def\Pi      {\ensuremath{\mathrm{i}}\xspace}

 \def\Pp      {\ensuremath{\mathrm{p}}\xspace}

 \def\Ps      {\ensuremath{\mathrm{s}}\xspace}

 \def\thebaroffset{0.0em}
}
{

 \def\Pmu         {\ensuremath{\mu}\xspace}

 \def\Ppi         {\ensuremath{\pi}\xspace}

 \def\Ppsi        {\ensuremath{\psi}\xspace}                 
                  
 \mathchardef\PDelta="7101
 \mathchardef\PXi="7104
 \mathchardef\PLambda="7103
 \mathchardef\PSigma="7106
 \mathchardef\POmega="710A
 \mathchardef\PUpsilon="7107
 \mathchardef\PPi="7105
                  
 \def\PB      {\ensuremath{B}\xspace}                 
                  
 \def\PD      {\ensuremath{D}\xspace}

 \def\PJ      {\ensuremath{J}\xspace}                 
 \def\PK      {\ensuremath{K}\xspace}

 \def\Pb      {\ensuremath{b}\xspace}                 
 \def\Pc      {\ensuremath{c}\xspace}

 \def\Pi      {\ensuremath{i}\xspace}

 \def\Pp      {\ensuremath{p}\xspace}

 \def\Ps      {\ensuremath{s}\xspace}

 \def\thebaroffset{0.18em}
}
\newcommand{\offsetoverline}[2][\thebaroffset]{\kern #1\overline{\kern -#1 #2}}%

\makeatletter
\ifcase \@ptsize \relax
  \newcommand{\miniscule}{\@setfontsize\miniscule{4}{5}}
\or
  \newcommand{\miniscule}{\@setfontsize\miniscule{5}{6}}
\or
  \newcommand{\miniscule}{\@setfontsize\miniscule{5}{6}}
\fi
\makeatother

\DeclareRobustCommand{\optbar}[1]{\shortstack{{\miniscule (\rule[.5ex]{1.25em}{.18mm})}
  \\ [-.7ex] $#1$}}

\def\mup        {{\ensuremath{\Pmu^+}}\xspace}
\def\mun        {{\ensuremath{\Pmu^-}}\xspace} 

\def\squark    {{\ensuremath{\Ps}}\xspace}

\def\cquark    {{\ensuremath{\Pc}}\xspace}

\def\bquark    {{\ensuremath{\Pb}}\xspace}

\def\pion   {{\ensuremath{\Ppi}}\xspace}

\def\pip    {{\ensuremath{\pion^+}}\xspace}
\def\pim    {{\ensuremath{\pion^-}}\xspace}

\def\kaon    {{\ensuremath{\PK}}\xspace}

\def\KorKbar {\kern \thebaroffset\optbar{\kern -\thebaroffset \PK}{}\xspace}

\def\Kp      {{\ensuremath{\kaon^+}}\xspace}
\def\Km      {{\ensuremath{\kaon^-}}\xspace}

\def\D       {{\ensuremath{\PD}}\xspace}

\def\DorDbar {\kern \thebaroffset\optbar{\kern -\thebaroffset \PD}\xspace}
\def\Dz      {{\ensuremath{\D^0}}\xspace}

\def\Dp      {{\ensuremath{\D^+}}\xspace}
\def\Dm      {{\ensuremath{\D^-}}\xspace}

\def\DpDm    {\ensuremath{\Dp {\kern -0.16em \Dm}}\xspace}

\def\Dsp     {{\ensuremath{\D^+_\squark}}\xspace}
\def\Dsm     {{\ensuremath{\D^-_\squark}}\xspace}

\def\B       {{\ensuremath{\PB}}\xspace}

\def\BorBbar {\kern \thebaroffset\optbar{\kern -\thebaroffset \PB}\xspace}

\def\Bd      {{\ensuremath{\B^0}}\xspace}

\def\BdorBdbar {\kern \thebaroffset\optbar{\kern -\thebaroffset \Bd}\xspace}
\def\Bu      {{\ensuremath{\B^+}}\xspace}

\def\Bp      {{\ensuremath{\Bu}}\xspace}

\def\Bs      {{\ensuremath{\B^0_\squark}}\xspace}

\def\BsorBsbar {\kern \thebaroffset\optbar{\kern -\thebaroffset \Bs}\xspace}

\def\jpsi     {{\ensuremath{{\PJ\mskip -3mu/\mskip -2mu\Ppsi}}}\xspace}

\def\Y#1S{\ensuremath{\PUpsilon{(#1S)}}\xspace}

\def\proton      {{\ensuremath{\Pp}}\xspace}

\def\Lz          {{\ensuremath{\PLambda}}\xspace}

\def\LorLbar     {\kern \thebaroffset\optbar{\kern -\thebaroffset \PLambda}\xspace}

\def\Xires       {{\ensuremath{\PXi}}\xspace}

\def\Lc          {{\ensuremath{\Lz^+_\cquark}}\xspace}

\def\Xic         {{\ensuremath{\Xires_\cquark}}\xspace}
\def\Xicz        {{\ensuremath{\Xires^0_\cquark}}\xspace}
\def\Xicp        {{\ensuremath{\Xires^+_\cquark}}\xspace}

\def\Lb           {{\ensuremath{\Lz^0_\bquark}}\xspace}

\def\Xibz         {{\ensuremath{\Xires^0_\bquark}}\xspace}
\def\Xibm         {{\ensuremath{\Xires^-_\bquark}}\xspace}

\def\BF         {{\ensuremath{\mathcal{B}}}\xspace}

\def\to                 {\ensuremath{\rightarrow}\xspace}

\def\AT#1     {\ensuremath{A_{\mathrm{T}}^{#1}}\xspace}           

\def\C#1      {\ensuremath{\mathcal{C}_{#1}}\xspace}                       
\def\Cp#1     {\ensuremath{\mathcal{C}_{#1}^{'}}\xspace}                    
\def\Ceff#1   {\ensuremath{\mathcal{C}_{#1}^{\mathrm{(eff)}}}\xspace}        
\def\Cpeff#1  {\ensuremath{\mathcal{C}_{#1}^{'\mathrm{(eff)}}}\xspace}       
\def\Ope#1    {\ensuremath{\mathcal{O}_{#1}}\xspace}                       
\def\Opep#1   {\ensuremath{\mathcal{O}_{#1}^{'}}\xspace}                    

\newcommand{\nospaceunit}[1]{\ensuremath{\text{#1}}}       
\newcommand{\aunit}[1]{\ensuremath{\text{\,#1}}}       

\newcommand{\tev}{\aunit{Te\kern -0.1em V}\xspace}
\newcommand{\gev}{\aunit{Ge\kern -0.1em V}\xspace}
\newcommand{\mev}{\aunit{Me\kern -0.1em V}\xspace}
\newcommand{\kev}{\aunit{ke\kern -0.1em V}\xspace}
\newcommand{\ev}{\aunit{e\kern -0.1em V}\xspace}
 
\newcommand{\mevc}{\ensuremath{\aunit{Me\kern -0.1em V\!/}c}\xspace}
\newcommand{\gevc}{\ensuremath{\aunit{Ge\kern -0.1em V\!/}c}\xspace}
\newcommand{\mevcc}{\ensuremath{\aunit{Me\kern -0.1em V\!/}c^2}\xspace}
\newcommand{\gevcc}{\ensuremath{\aunit{Ge\kern -0.1em V\!/}c^2}\xspace}

\def\mum  {\ensuremath{\,\upmu\nospaceunit{m}}\xspace}

\def\fb   {\ensuremath{\aunit{fb}}\xspace}
\def\invfb   {\ensuremath{\fb^{-1}}\xspace}

\def\gsim{{~\raise.15em\hbox{$>$}\kern-.85em
          \lower.35em\hbox{$\sim$}~}\xspace}
\def\lsim{{~\raise.15em\hbox{$<$}\kern-.85em
          \lower.35em\hbox{$\sim$}~}\xspace}

\def\sqs   {\ensuremath{\protect\sqrt{s}}\xspace}

\def\pt         {\ensuremath{p_{\mathrm{T}}}\xspace}

\def\evtgen     {\mbox{\textsc{EvtGen}}\xspace}

\def\geant      {\mbox{\textsc{Geant4}}\xspace}

\def\photos     {\mbox{\textsc{Photos}}\xspace}

\def\pythia     {\mbox{\textsc{Pythia}}\xspace}

\def\tell1  {TELL1\xspace}
\def\ukl1   {UKL1\xspace}

\newcommand{\ie}{\mbox{\itshape i.e.}\xspace}

\newcommand{\lhcborcid}[1]{\href{https://orcid.org/#1}{\hspace*{0.1em}\raisebox{-0.45ex}{\includegraphics[width=1em]{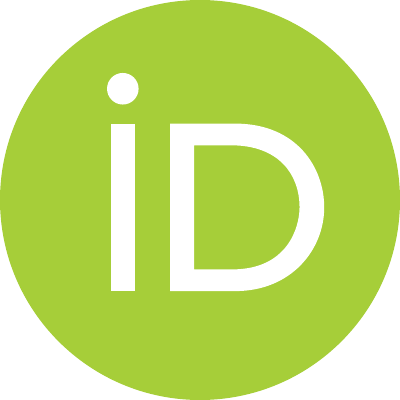}}}}

\usepackage{cite} 
\usepackage{mciteplus}

\usepackage{longtable} 
\usepackage{tabularx}

\begin{document}

\renewcommand{\thefootnote}{\fnsymbol{footnote}}
\setcounter{footnote}{1}


\begin{titlepage}
\pagenumbering{roman}

\vspace*{-1.5cm}
\centerline{\large EUROPEAN ORGANIZATION FOR NUCLEAR RESEARCH (CERN)}
\vspace*{1.5cm}
\noindent
\begin{tabular*}{\linewidth}{lc@{\extracolsep{\fill}}r@{\extracolsep{0pt}}}
\ifthenelse{\boolean{pdflatex}}
{\vspace*{-1.5cm}\mbox{\!\!\!\includegraphics[width=.14\textwidth]{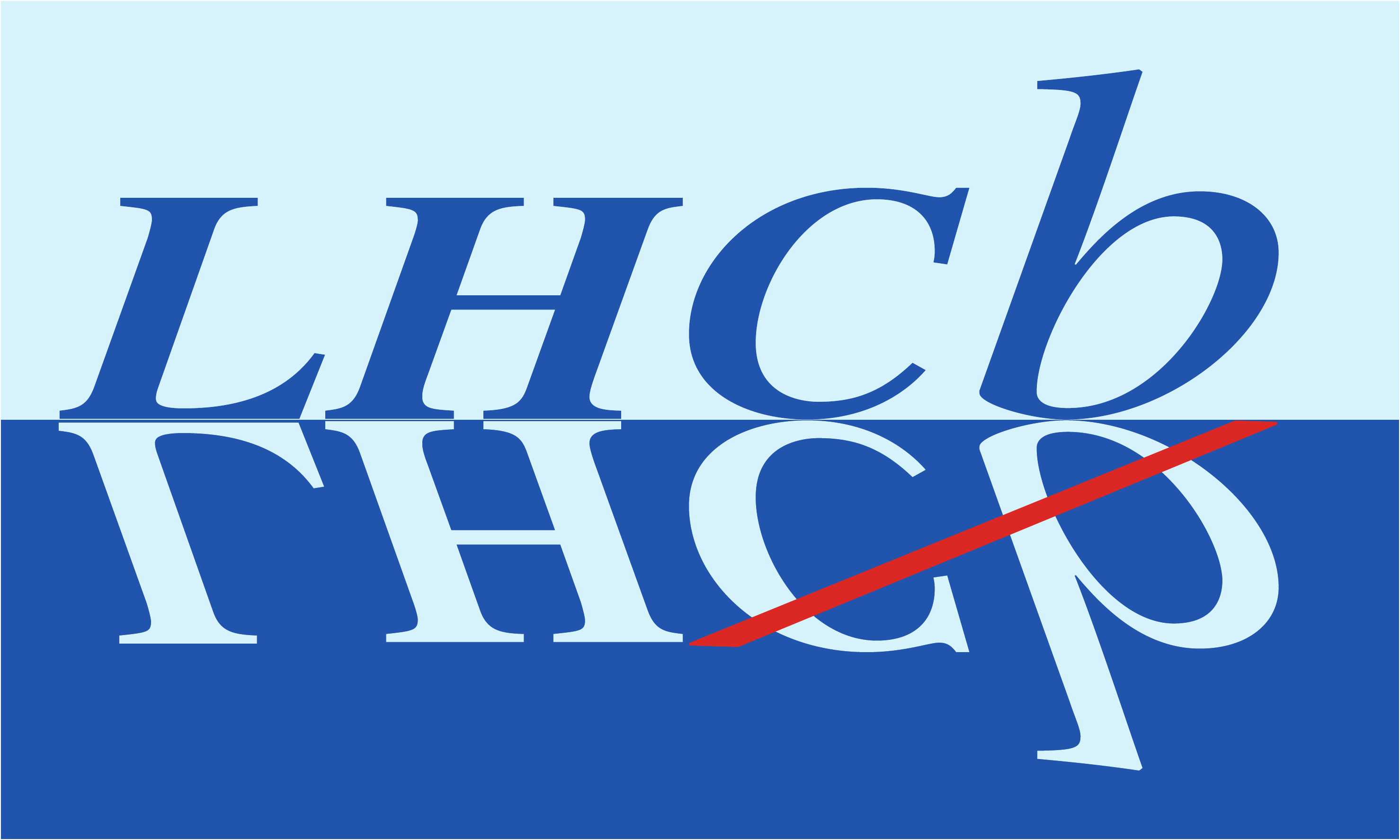}} & &}%
{\vspace*{-1.2cm}\mbox{\!\!\!\includegraphics[width=.12\textwidth]{figs/lhcb-logo.eps}} & &}%
\\
 & & CERN-EP-2023-173 \\  
 & & LHCb-PAPER-2023-017 \\  
 & & October 20, 2023 \\
 & & \\
\end{tabular*}

\vspace*{2.0cm}

{\normalfont\bfseries\boldmath\huge
\begin{center}
  \papertitle 
\end{center}
}

\vspace*{1.0cm}

\begin{center}
\paperauthors\footnote{Authors are listed at the end of this paper.}
\end{center}

\vspace{\fill}

\begin{abstract}
  \noindent
The  $\Xibz \to \Xicp \Dsm$ and
$\Xibm \to \Xicz \Dsm$ decays are observed for the first time using proton-proton collision data collected by the LHCb experiment at a centre-of-mass energy of $\sqs=13\tev$,
corresponding to an integrated luminosity of $5.1\invfb$.
The relative branching fractions times the beauty-baryon production cross-sections are measured to be
  \begin{align*}
    \mathcal{R}\left(\frac{\Xibz}{\Lb}\right) &\equiv
    \frac{\sigma\left(\Xibz\right)}{\sigma\left(\Lb\right)}
    \times
    \frac{\BF\left(\Xibz \to \Xicp \Dsm\right)}{\BF\left(\Lb \to \Lc \Dsm\right)}
    =(15.8\pm1.1\pm0.6\pm7.7)\%,\\
    \mathcal{R}\left(\frac{\Xibm}{\Lb}\right) &\equiv
    \frac{\sigma\left(\Xibm\right)}{\sigma\left(\Lb\right)}
    \times
    \frac{\BF\left(\Xibm \to \Xicz \Dsm\right)}{\BF\left(\Lb \to \Lc \Dsm\right)}
    =(16.9\pm1.3\pm0.9\pm4.3)\%,
  \end{align*}
  where the first uncertainties are statistical, the second systematic, 
  and the third due to the uncertainties on the branching fractions of relevant charm-baryon decays.
The masses of \Xibz and \Xibm baryons are measured to be \mbox{$m_{\Xibz}=5791.12\pm0.60\pm0.45\pm0.24\mevcc$} and \mbox{$m_{\Xibm}=5797.02\pm0.63\pm0.49\pm0.29\mevcc$},
where the uncertainties are statistical, systematic, 
  and those due to charm-hadron masses, respectively.
  
\end{abstract}


\begin{center}
  Submitted to
  Eur.~Phys.~J.~C 
\end{center}

\vspace{\fill}

{\footnotesize 
\centerline{\copyright~\papercopyright. \href{\paperlicenceurl}{\paperlicence}.}}
\vspace*{2mm}

\end{titlepage}


\newpage
\setcounter{page}{2}
\mbox{~}
%
%
%
%


\renewcommand{\thefootnote}{\arabic{footnote}}
\setcounter{footnote}{0}

\cleardoublepage


\pagestyle{plain} 
\setcounter{page}{1}
\pagenumbering{arabic}


\section{Introduction}
\label{sec:Introduction}

Hadrons are systems of quarks bound by the strong interaction, described at the fundamental level by quantum chromodynamics (QCD). The production and decay of hadrons involve the nonperturbative regime of QCD, making calculations challenging. Much progress has been made in recent years in experimental and theoretical studies of beauty mesons, with the aim of testing the Standard Model and searching for new physics through measurements of branching fractions, $CP$ asymmetries and rare decays~\cite{Chen:2021ftn}.  
However, many aspects of beauty baryons are still largely unknown, due to the difficulties to produce and detect them in experiments other than those operating at the Large Hadron Collider.

So far, the \Lb baryon has been more widely studied than the other beauty baryons, including \Xibz and \Xibm.\footnote{The inclusion of charge-conjugate processes is implied throughout.}
Very few decay modes have been measured for $\Xires_b^{0(-)}$ baryons~\cite{PDG2022}.
According to the quark model, the three beauty baryons $\Lb$, $\Xibz$ and $\Xibm$ (referred to as $H_b$ in the following) form an $SU(3)$ flavour multiplet, as do the $\Lc$, $\Xicp$ and $\Xicz$ states (referred to as $H_c$ in the following). 
The $H_b$ decay is dominated by the weak transition of the $b$ quark while the two light quarks serve as compact spectators~\cite{Lenz:2014jha,Neubert:1997gu}.  
According to heavy quark effective theory,
the three decays of bottom baryons into two charm hadrons, $H_b\to H_c \Dsm$, should have approximately the same partial width~\cite{Neubert:1993mb,Chua:2019yqh}. 
The $\Lb\to \Lc\Dsm$ decay has been measured to have a branching fraction ($\BF$) at the percent level~\cite{LHCb-PAPER-2014-002}, but no measurements for $\Xires_b^{0(-)}\to \Xires_c^{+(0)}\Dsm$ decays are available. Measurements of these decays
not only test the $SU(3)$ symmetry but also give insights into the dynamics of beauty-baryon weak decays. 

Beauty baryons of all species are abundantly produced at the LHC ~\cite{LHCB-PAPER-2011-018,LHCB-PAPER-2014-004,LHCb-PAPER-2018-050,LHCb-PAPER-2018-047}, allowing them to be intensively studied.
This analysis presents the first observation of 
$\Xibz \to \Xicp \Dsm$ and 
$\Xibm \to \Xicz \Dsm$ decays, using data from proton-proton ($pp$) collisions at a centre-of-mass energy of $\sqs=13\tev$ collected by \lhcb detector and corresponding to an integrated luminosity of $5.1\invfb$. 
The relative production rates of the decays, $\mathcal{R}$, defined to be
\begin{align}
  \mathcal{R}\left(\frac{\Xibz}{\Lb}\right) &\equiv 
  \frac{\sigma\left(\Xibz\right)}{\sigma\left(\Lb\right)} \times \frac{\BF\left(\Xibz \to \Xicp \Dsm\right)}{\BF\left(\Lb \to \Lc \Dsm\right)},\label{eq:R_xibz_lb}\\
  \mathcal{R}\left(\frac{\Xibm}{\Lb}\right) &\equiv 
  \frac{\sigma\left(\Xibm\right)}{\sigma\left(\Lb\right)} \times \frac{\BF\left(\Xibm \to \Xicz \Dsm\right)}{\BF\left(\Lb \to \Lc \Dsm\right)},\label{eq:R_xibm_lb}\\
  \mathcal{R}\left(\frac{\Xibz}{\Xibm}\right) &\equiv 
  \frac{\sigma\left(\Xibz\right)}{\sigma\left(\Xibm\right)} \times \frac{\BF\left(\Xibz \to \Xicp \Dsm\right)}{\BF\left(\Xibm \to \Xicz \Dsm\right)}, \label{eq:R_xibz_xibm}
\end{align}
are measured,
where $\sigma$ denotes the production cross-section.
Given the similar lifetimes of the three beauty baryons~\cite{PDG2022}, if the decay widths of the three beauty-baryon decays are also similar, the variables defined in Eq.~\ref{eq:R_xibz_lb}-~\ref{eq:R_xibz_xibm} provide measurements of the $H_b$ production cross-section ratios, \ie $b$-quark fragmentation fraction ratios. 
Isospin symmetry assures that  $\sigma\left(\Xibz\right)/\sigma\left(\Xibm\right)\approx1$ to a good approximation, resulting in $\mathcal{R}\left(\frac{\Xibz}{\Xibm}\right)\approx1$ at leading order, which is tested in this analysis.
The masses of the \Xibz and \Xibm baryons and the mass differences between the three beauty baryons are also measured.

\section{Detector, samples and analysis strategy}%
\label{sec:detector_simulation}

The LHCb detector~\cite{LHCb-DP-2014-002, LHCb-DP-2008-001} is a single-arm forward spectrometer covering the pseudorapidity range $2<\eta<5$, designed for the study of particles containing $b$ or $c$ quarks. 
The detector includes a high-precision tracking system consisting of a silicon-strip vertex detector surrounding the $pp$ interaction region, a large-area silicon-strip detector located upstream of a dipole magnet with a bending power of about 4 Tm, and three stations of silicon-strip detectors and straw drift tubes placed downstream of the magnet. The tracking system provides a measurement of the momentum, $p$, of charged particles with a relative uncertainty that varies from 0.5\% at low momentum to 1.0\% at 200\gevc. 
The momentum scale is calibrated using samples of $\jpsi \to \mup \mun$ and $\Bp \to \jpsi \Kp$ decays collected concurrently with the data samples used for this analysis~\cite{LHCb:2013wmn, LHCb:2013fwo}. The relative uncertainty of this procedure is determined to be $3\times 10^{-4}$ using samples of other fully reconstructed $B$, $\Upsilon$, and $K_{S}^0$-meson decays.
The minimum distance of a track to a primary $pp$ collision vertex (PV), the impact parameter (IP), is measured with a resolution of $(15+29/\pt)$\mum, where $\pt$ is the component of the momentum transverse to the beam, in \gevc. Different types of charged hadrons are distinguished using information from two ring-imaging Cherenkov detectors. Photons, electrons and hadrons are identified by a calorimeter system consisting of scintillating-pad and preshower detectors, an electromagnetic and a hadronic calorimeter. 

The data used in this analysis come from $pp$ collisions at 
$\sqs=13\tev$, 
collected by LHCb between 2016 and 2018. The total integrated luminosity is $5.1\invfb$. 
The online event selection of \lhcb is performed by a trigger~\cite{LHCb-DP-2019-001}, which consists of a hardware stage, based on information from the calorimeter and muon systems, followed by a software stage, which applies a full event reconstruction. At the hardware trigger stage, events are required to have a muon with high $\pt$ or a hadron, photon or electron with high transverse energy in the calorimeters. 
A global hardware trigger decision is required based on the reconstructed candidate, the rest of the event, or a combination of both.
The software trigger requires a two-, three- or four-track secondary vertex with a significant displacement from any primary $pp$ interaction vertex. At least one charged particle within the secondary vertex must have a transverse momentum $\pt>1.6\gevc$ and be inconsistent with originating from any PV.

Simulated decays are used to perform event selections, calculate reconstruction and selection efficiencies, and determine the invariant-mass distributions of the reconstructed signal $H_b$ candidates. In the simulation, $pp$ collisions are generated using \pythia8~\cite{Sjostrand:2007gs} with a specific LHCb configuration~\cite{LHCb-DP-2008-001}. Decays of unstable particles are described by \evtgen ~\cite{Lange:2001uf}, in which final-state radiation is generated using \photos ~\cite{davidson2015photos}. 
The interaction of the generated particles with the detector, and its response, are simulated using the \geant ~\cite{Allison:2006ve} toolkit as described in Ref.~\cite{Clemencic_2011}.

The $\Lc$ and $\Xicp$ baryons are reconstructed in the $p\Km\pip$ final state, 
and the $\Xicz$ baryon in the $p\Km\Km\pip$ final state. The $\Dsm$ mesons are reconstructed by combining three charged particles identified as $\Km$, $K^+$ and $ \pi^-$ mesons. The $H_c$ candidates are combined with $\Dsm$ candidates to form the $H_b$ candidates. 
The three $\mathcal{R}$ parameters are defined according to 
\begin{align}
  \mathcal{R} \left( \frac{\Xibz}{\Lb}\right)& = \frac{ N \left( \Xibz \to \Xicp \Dsm\right) / \varepsilon \left( \Xibz \to \Xicp \Dsm \right)}{N \left( \Lb \to \Lc \Dsm\right) / \varepsilon\left(\Lb \to \Lc \Dsm\right)}
  \times 
  \frac{\BF\left(\Lc \to p \Km \pip \right)}
  {\BF \left( \Xicp \to p \Km \pip\right)},\\
  \mathcal{R}\left(\frac{\Xibm}{\Lb}\right)&=\frac{N\left(\Xibm \to \Xicz \Dsm\right) / \varepsilon\left(\Xibm \to \Xicz \Dsm\right)}
  {N\left(\Lb \to \Lc \Dsm\right) / \varepsilon\left(\Lb \to \Lc \Dsm\right)}
  \times
  \frac{\BF\left(\Lc \to p \Km \pip\right)}
  {\BF\left(\Xicz \to p \Km \Km \pip\right)},\\
  \mathcal{R}\left(\frac{\Xibz}{\Xibm}\right)&=\frac{N\left(\Xibz \to \Xicp \Dsm\right) / \varepsilon\left(\Xibz \to \Xicp \Dsm\right)}
  {N\left(\Xibm \to \Xicz \Dsm\right) / \varepsilon\left(\Xibm \to \Xicz \Dsm\right)}
  \times
  \frac{\BF\left(\Xicz \to p \Km \Km \pip\right)}
  {\BF\left(\Xicp \to p \Km \pip\right)},
\end{align}
where $N$, $\varepsilon$, and \BF denote the observed signal yields, the total experimental efficiencies, and the branching fractions, respectively. 
The world averages of branching fractions of corresponding $H_c$ decays~\cite{PDG2022} are summarised in Table~\ref{tab:Hc_branching_fraction}.
The signal yields are determined using unbinned extended maximum-likelihood fits of the $H_c\Dsm$ invariant-mass distributions. The efficiencies are determined using simulated signal decays, calibrated by data driven methods.

\begin{table}[tb]
  \centering
  \caption{Branching fractions of $H_c$ decays~\cite{PDG2022}.}
  \label{tab:Hc_branching_fraction}
  \begin{tabular}{l c}
    \hline
    Decay & Branching fraction\\
    \hline
    $\Lc \to p\Km \pip$ & $(6.28\pm0.32)\times10^{-2}$\\
    $\Xicp \to p \Km \pip$ & $(6.2\pm3.0)\times10^{-3}$\\
    $\Xicz \to p \Km \Km \pip$ & $(4.8\pm1.2)\times10^{-3}$\\
    \hline
  \end{tabular}
\end{table}

\section{Event selections and efficiencies}%
\label{sec:selection}
In order to suppress background due to random combinations of either the $H_c$ or \Dsm, and misidentification of final-state particles, a series of event selections are performed. Firstly, all final-state particles are required to be separated from any PV and have $\pt>100\mevc$. They must also be correctly identified, with a high significance, as either a proton, kaon or pion, using combined information from the tracking system and sub-detectors related to particle identification (PID)~\cite{LHCb-DP-2014-002,LHCb-DP-2018-001}. The final states of the $H_c$ and \Dsm candidates must have a scalar sum of $\pt > 1.8\gevc$, and at least one of them must have $\pt>0.5\gevc$ and $p>5\gevc$. They are additionally required to form a good vertex that is significantly separated from any PV. The $H_c$ and \Dsm candidates should have an invariant mass within $\pm25\mevcc$ of the previous world average mass value~\cite{PDG2022}, and their vertices should be consistent with being downstream of the $H_b$ vertex.
The $H_b$ candidate formed by the $H_c$ and $\Dsm$ hadrons must have a good vertex separated from its associated PV, and its momentum must point back to the associated PV. The final-state particles of the $H_b$ must have a scalar sum of $\pt>5\gevc$. Finally, $H_b$ candidates with transverse momentum $\pt>4\gev$ and rapidity $2.5<y<4$ are retained for further analysis.

There are backgrounds due to genuine particle decays, where a pion or kaon decay product is misidentified as a proton, resulting in a $H_c$ candidate.
 For $\Lc$ and $\Xicp$ candidates, they include $\phi\to \Kp\Km$, $\Dsp\to\Kp\Km\pip$, $\Dp\to\Kp\Km\pip$ and $\Dz\to\Kp\Km$ decays with the $\Kp$ meson misidentified as a proton,  and $\Dp\to\Km\pip\pip$, $\Dz\to\Km\pip$ decays with the $\pip$ meson misidentified as a proton. For $\Xicz$ candidates, there are backgrounds due to $\phi\to \Kp\Km$ and $\Dz\to\Kp\Km\Km\pip$ decays with the $\Kp$ meson misidentified as a proton. 
For $\Dsm$ candidates, the $\Lc\to\proton\Km\pip$ background with the proton misidentified as a $\Kp$ meson is considered. 
To remove these background, candidates are required to satisfy strict PID requirements or their invariant masses, calculated with alternative mass hypotheses for final states, must be outside a region around the known mass of the corresponding genuine particle ($\phi$, \Dsp, $D^+$, $D^0$, or \Lc)~\cite{PDG2022}.
Backgrounds due to $D^- \to \Kp \pim\pim$ decays are also considered, and are found to be negligible.

Further event selections are performed using a gradient-boosted decision tree (BDTG)~\cite{AdaBoost} algorithm to reduce combinatorial backgrounds.
Due to the similarity between the topologies of the three $H_b\to H_c \Dsm$ decays, and to benefit from a cancellation of systematic uncertainties related to the BDTG selection in the $\mathcal{R}$ measurements, the BDTG classifier is trained with the \Xibz samples and is applied to all the three decay modes. 
The BDTG algorithm is trained to distinguish simulated $\Xibz \to \Xicp \Dsm$ decays from the candidates in the high mass sideband ($m(\Xibz)>5950\mevcc$) of data, which are representative of the background.
The BDTG classifier combines seventeen variables, including kinematic, topological and PID information, to get a single discriminating response. 
The optimal requirement on the BDTG response is determined by maximising the figure of merit $F\equiv S/\sqrt{S+B}$, where $S$ ($B$) is the expected number of signal (background) yield in the signal region of data with BDTG response greater than a given value.
The signal region is defined to be $\pm30\mevcc$ around the previous world average of $H_b$ mass~\cite{PDG2022}, which is about three times the experimental resolution.
The value of $S$ is calculated as the product of the BDTG efficiency for the signal and the signal yield before the BDTG requirement,  which is obtained by fitting to \Xibz invariant-mass distribution in data.
Similarly, $B$ is calculated as the background retention rate multiplied by the estimated background in the signal region without the BDTG requirement. The background retention rate is evaluated with the data in the high-mass sideband data, and the number of background candidates in the signal region is estimated with a fit to \Xibz invariant-mass distribution in the high invariant-mass sideband region of the data, with a subsequent extrapolation to the signal mass region.
The optimal BDTG requirement corresponds to a signal efficiency of about 95\% with respect to other selection requirements for all three $H_b$ decay modes.

The total efficiency is calculated as the product of efficiencies of detector acceptance, reconstruction, and selection. It is estimated using the simulated signal decays.
These samples are calibrated such that the shapes of several key distributions match those of the data: the PID response, $H_b$ kinematics, total charged-track multiplicity and $H_c$ resonant structures. The $\Dsm\to\Kp\Km\pim$ decay is simulated using measured Dalitz compositions~\cite{Dalitz:1953cp}, thus no corrections are applied. 
The PID efficiencies for the different particle species are measured using charm hadron samples in data~\cite{LHCb-DP-2018-001}.
The large sample of $\Lb\to\Lc\Dsm$ decays is used to correct for the transverse momentum, pseudorapidity, and charged-track multiplicity distributions of the three $H_b$ decay modes. 
Further corrections are made to align the shapes of the charged-track multiplicity distributions in the data and simulation for $\Xires_b$ decays. 
The $H_c$ Dalitz distribution is compared between the data and simulation; a weight-based correction is applied to improve the agreement.
The track-finding efficiency in simulation is found to be slightly different from that in data, and this difference is corrected as a function of the momentum and pseudorapidity of final-state particles~\cite{LHCb-DP-2013-002}.
The correction factors are generally obtained in bins of relevant variables apart from that for the $\Xicz$ Dalitz distribution, where the large number of dimensions implies a limited number of candidates per bin. An unbinned multivariate algorithm is therefore used~\cite{Rogozhnikov:2016bdp}.
The ratios of efficiencies between $\Lb$, $\Xibz$, and $\Xibm$ decays are determined to be
\begin{align*}
    \frac{\varepsilon(\Xibz)}{\varepsilon(\Lb)}=1.101\pm0.010,\\
    \frac{\varepsilon(\Xibm)}{\varepsilon(\Lb)}=0.515\pm0.005,\\
    \frac{\varepsilon(\Xibz)}{\varepsilon(\Xibm)}=2.138\pm0.017,
\end{align*}
where the uncertainties are statistical only. The $\Lb$ and $\Xibz$ decays have a similar efficiency, while the smaller $\Xibm$ efficiency is due to one more final-state particle.

\section{Signal yield determination and mass measurements}
\label{sec:fit}

To obtain the yields of signal $H_b$ decays, an extended maximum likelihood fit is performed to the $\Lb$, $\Xibz$, and $\Xibm$ invariant-mass spectra. 
A kinematic refit~\cite{Hulsbergen:2005pu} is  applied to the $H_b$ decays to improve the mass resolution, constraining the $\Dsm$ and $H_c$ masses to their previously measured values~\cite{PDG2022} and the $H_b$ momentum to point back to its PV.
The fitted mass region is $5450$ -- $5800\mevcc$, $5600$ -- $6100\mevcc$, and $5600$ -- $6000\mevcc$ for the $\Lb$, $\Xibz$, and $\Xibm$ decays, respectively. 

\begin{figure}[!tb]
  \centering
  \includegraphics[width=0.49\linewidth]{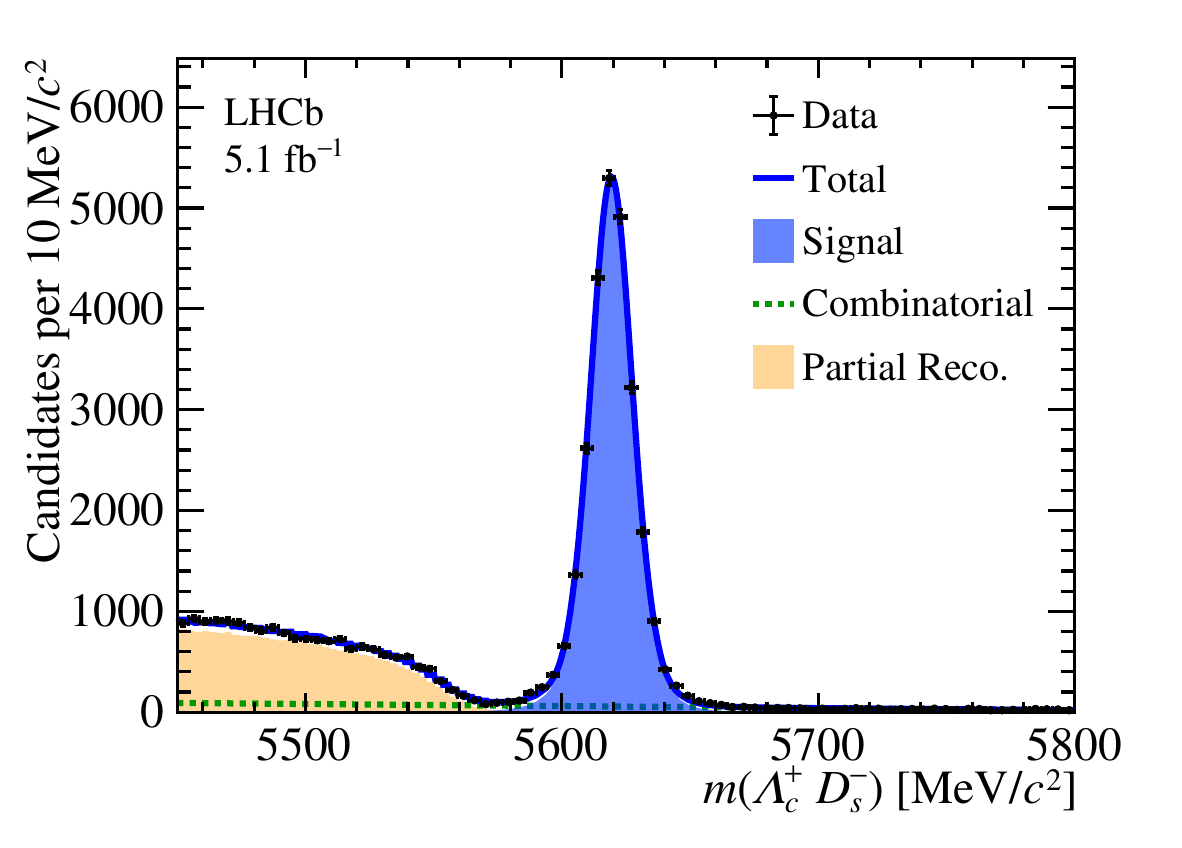}
  \includegraphics[width=0.49\linewidth]{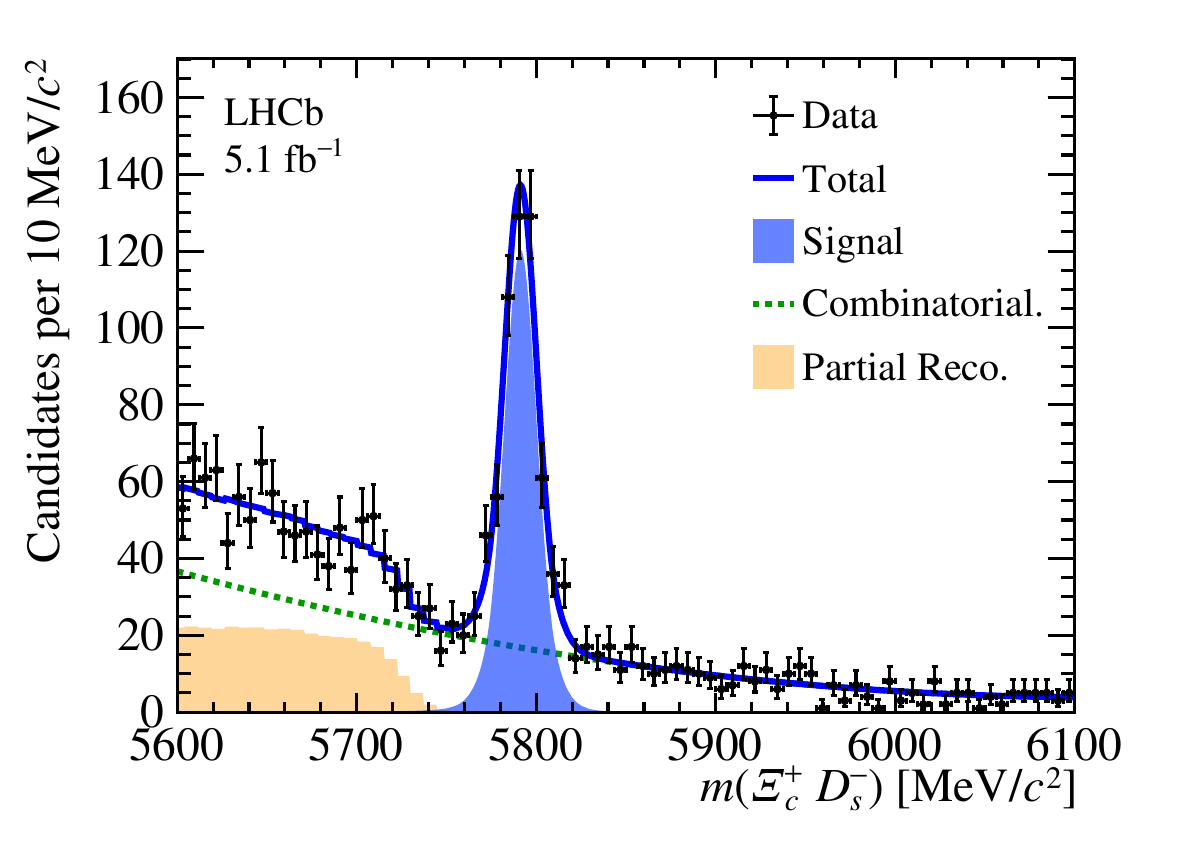}
  \includegraphics[width=0.49\linewidth]{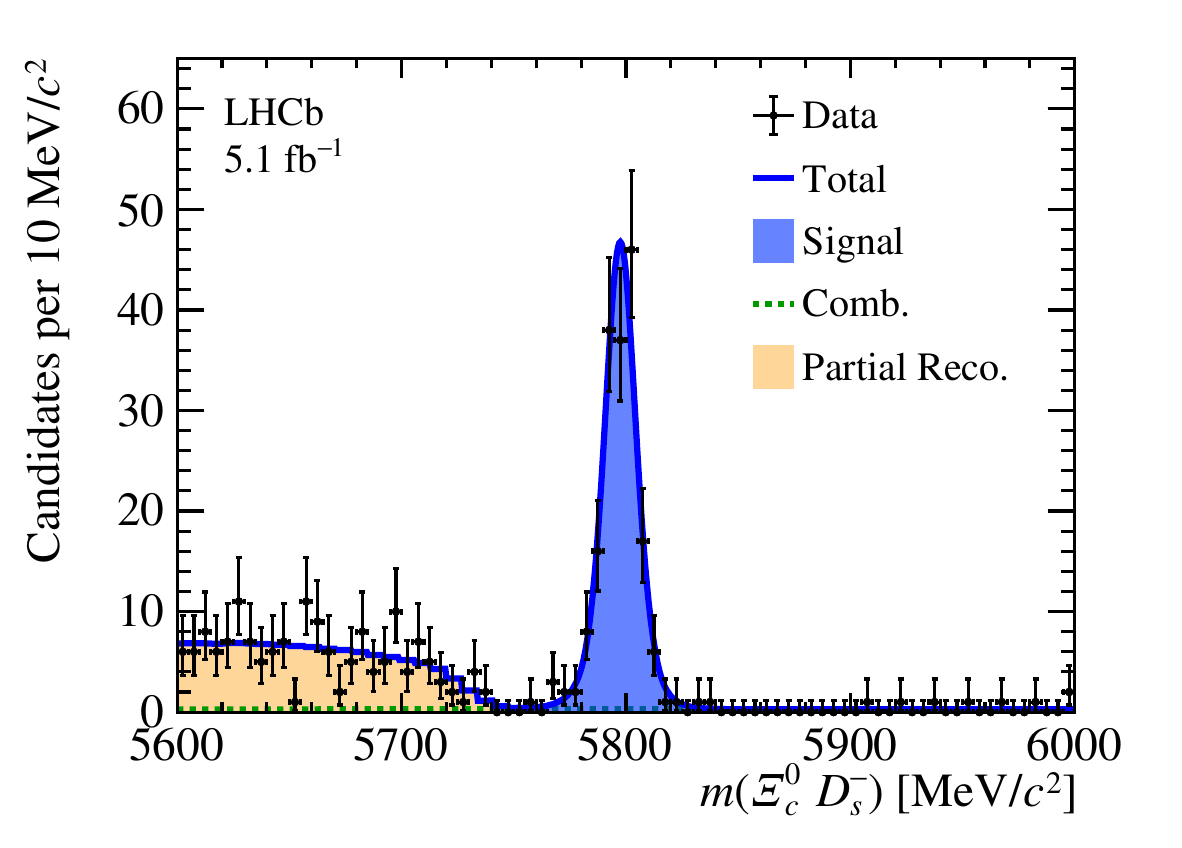}
  \caption{Invariant-mass distributions of (top left) $\Lb$ , (top right) $\Xibz$ , and (bottom)  $\Xibm$ decays. The data are overlaid on the fit results.}%
  \label{fig:fit}
\end{figure}

As shown in Fig.~\ref{fig:fit}, three components are identified in each $H_b$ mass spectrum.
The signal component is parameterised using the sum of a Gaussian and a double-sided Crystal Ball function (DSCB)~\cite{Skwarnicki:1986xj} sharing a common mean.
The common mean and the average resolution of the Gaussian and the DSCB distribution are parameters that vary freely in the fit, while the other parameters have values fixed to those obtained from simulation.
The contribution of combinatorial backgrounds in the mass spectrum is modelled using a second order polynomial, with all parameters varying freely.
The peaking structure in the low invariant-mass region  corresponds to partially reconstructed $H_b\to H_c\Dsm X$ decays where $X$ is an undetected particle. Distributions from data in the low mass region are found to be consistent with the $H_b\to H_c D_s^{*-}$, $D_s^{*-}\to\Dsm\gamma$ sequential decay, where the $\gamma$ is not reconstructed. The subsequent $H_c\Dsm$ invariant-mass distribution depends on the $D_s^{*-}$ helicity projection, for which three possibilities, helicities of $\pm1$ and 0, are allowed. The mass distributions for helicities of $+1$ and $-1$ are identical. Samples are generated with helicities of $1$ and $0$, and corresponding $H_c\Dsm$ invariant-mass distributions are obtained. The distributions convoluted with experimental resolutions are used to fit data. The fraction of the component with a helicity of 0 varies freely in the fit.


Figure~\ref{fig:fit} shows the $H_b$ invariant-mass distributions superimposed by the fit results. 
The signal yields for $\Lb$, $\Xibz$ and $\Xibm$ decays are $(2.609\pm0.017)\times10^4 $, $462\pm29$, and $175\pm14$, respectively.
The masses for $\Lb$, $\Xibz$ and $\Xibm$ baryons are measured to be \mbox{$m_{\Lb}=5619.34\pm0.06\mevcc$}, \mbox{$m_{\Xibz}=5791.12\pm0.60\mevcc$}, and \mbox{$m_{\Xibm}=5797.02\pm0.63\mevcc$}, respectively, where the uncertainties are statistical only.

\subsection{Non-dicharm background}%
\label{sec:charmless_bkg}
The sample of $H_b \to H_c D_s$ decays is polluted by decays with a single charm hadron (one-charm) or charmless decays that have the same final-state particles but without the intermediate $H_c$ or $\Dsm$ hadron, referred to as non-dicharm decays. 
For these peaking background contributions, their $H_b$ invariant-mass distributions are signal-like, but the $H_c$ or $\Dsm$ invariant-mass distributions are flat.
The  distributions of non-dicharm components in the $H_c$ or $\Dsm$ invariant-mass distribution are found to be approximately linear. Therefore, the $H_b$ signal yields in the $H_c$ and $\Dsm$ sideband regions are extrapolated to the signal region ($\pm25\mevcc$ around the previously measured $H_c$ and $\Dsm$ masses~\cite{PDG2022}) to estimate the contamination of non-dicharm background in the signal region.
Details of the estimation are shown in 
Appendix~\ref{sec:non-dicharm}.
The fractions of non-dicharm decays are measured to be $(5.70\pm0.13)\%$, $(8.39\pm1.75)\%$ and $(6.44\pm1.48)\%$ for $\Lb$, $\Xibz$ and $\Xibm$ decays, respectively. These background contributions are subtracted from the total signal yield obtained from the fit. The non-dicharm contamination is dominated by the $H_b\to H_c(\Kp\Km\pim)$ component.

\section{Systematic uncertainties}%
\label{sec:systematic_uncertainties}

\subsection{Uncertainties on the branching fraction}

Measurements of the ratios of branching fractions 
are affected by a number of systematic uncertainties. Apart from those due to the input charm-decay branching fractions, they are generally related to either the signal yields or the efficiencies.
Due to the similar topologies of the three $H_b$ decays, many sources of systematic uncertainties are either cancelled or largely suppressed in ratios of the branching fractions. 
The remaining systematic uncertainties are outlined below and summarised in Table~\ref{tab:uncertain}.

\subsubsection{Systematic uncertainties on the signal yield}
\label{sec:uncertain_mass_fit}

The fit results are affected by the imperfect modelling of the signal, the combinatorial background and the partially reconstructed background.
Variations of the signal model are studied by modifying the fixed parameters that are obtained from simulation. 
For the background modelling, a polynomial of third order is used instead of one of second order.
In order to study the impact of the modelling of the partially reconstructed background in the signal yield, 
the lower edge of the fit range is increased to 5575, 5740, and 5750$\mevcc$ for the $\Lb$, $\Xibz$, and $\Xibm$ decay modes, respectively, excluding partially reconstructed background.
Alternative fits to data with these alternate approaches are performed.
The largest deviation of the $H_b$ signal yield in these alternative fits from the nominal result 
is taken as the systematic uncertainty on the signal yield due to the modelling of the fit components, which is at the level of 2\%.

The uncertainty on the fraction of non-dicharm background discussed in Section~\ref{sec:charmless_bkg} originates from the limited size of the data sample and possible nonlinearity of the $H_c$ and $\Dsm$ background invariant-mass distributions. The effect is studied by using alternative regions of sideband data to calculate the non-dicharm yield, and the difference with respect to the nominal results is quoted as the systematic uncertainty, which is found to be at the subpercent level.

\subsubsection{Systematic uncertainties on the efficiency}

As efficiencies are studied using simulation samples, 
the systematic uncertainty on efficiencies arises due to the limited size of simulation samples and imperfect simulations.
The uncertainty due to the limited simulation sample size is 1.0\% for the three $H_b$ efficiency ratios.

The hardware trigger is approximately modeled in the simulation.
The trigger efficiency in measured in the data~\cite{LHCb-DP-2020-001}, and the difference between data and simulation is assigned as a systematic uncertainty. 
This systematic uncertainty is found to be approximately cancelled among the three $H_b$ decay modes, resulting in a relative difference of  less than 1.5\% between data and simulation on the efficiency ratios of the two $H_b$ decay modes.
A common value of 1.5\% is quoted as the relative systematic uncertainty of the hardware trigger on the relative branching fraction.

The estimation of the reconstruction efficiency is affected by the model of detector material in simulation which affects the description of interaction between the final-state particles and the material. It leads to a relative uncertainty of 1.2\% between $\Xibm$ and the other two $H_b$ decays due to one additional kaon in the $\Xibm$ decay~\cite{LHCb:2014nio}.
Moreover, the estimation of the track-finding efficiency in data and simulation is subjected to uncertainties related to the detector occupancy and limited sizes of the calibration samples~\cite{LHCb-DP-2013-002}. The former gives a relative value of 0.8\%  per track, while the latter results in an uncertainty of around 0.1\% on the efficiency ratios. In total the uncertainty on the ratio of reconstruction efficiency is about 1.6\% between $\Xibm$ and $\Lb$ decays, and between $\Xibz$ and $\Xibm$ decays. It is below $0.1\%$ for the efficiency ratio between $\Xibz$ and $\Lb$ decays.

Corrections to simulation samples to match data to the distributions of final-state particle PID responses, $H_b$ kinematics, charged-track multiplicity and $H_c$ Dalitz distributions are subject to uncertainties. Uncertainties on the corrections of PID responses are evaluated using alternative corrections and measuring the relative change of efficiencies~\cite{LHCb-DP-2018-001}, which is found to be negligible. The uncertainty on corrections of $H_b$ kinematics is studied with pseudoexperiments. For each pesudoexperiment, the correction factor in each transverse momentum and rapidity of the $H_b$ baryon is varied following a Gaussian distribution constructed from the nominal value and its uncertainty. The new correction factors are used to calculate the efficiency. The width of the efficiency distribution among a set of pseudoexperiments is taken as the systematic uncertainty. Similar studies are performed for corrections of the charge-track multiplicity and $\Lc$, $\Xicp$ Dalitz distributions. The uncertainty of the unbinned correction to the $\Xicz$ Dalitz distribution is studied by varying the configurations of the algorithm~\cite{Rogozhnikov:2016bdp}. In total the uncertainty on the efficiency ratio originating from corrections to simulation samples is about 4.3\% between $\Xibm$ and $\Lb$, 4.3\% between $\Xibz$ and $\Xibm$, and 1.3\% between $\Xibz$ and $\Lb$.

\begin{table}[tb]
    \centering
    \caption{Systematic uncertainties on the  relative branching fraction measurements. Results are given as relative uncertainties.}
    \begin{tabular}{l c c c}
        \hline \\ [-1.5ex]
        Source & $\mathcal{R}\left(\frac{\Xibz}{\Lb}\right)$ & $\mathcal{R}\left(\frac{\Xibm}{\Lb}\right)$ & $\mathcal{R}\left(\frac{\Xibz}{\Xibm}\right)$\\ [1.5ex]
        \hline
        
        Imperfect modelling of invariant-mass fit & 2.7\% & 1.3\% & 3.4\% \\
        Fraction of non-dicharm background & 2.0\% & 1.6\% & 2.5\% \\
        Limited simulation sample size &  0.9\% & 1.0\% & 0.8\% \\
        Trigger efficiency & 1.5\% & 1.5\% & 1.5\% \\
        Reconstruction efficiency & 0.1\% & 1.6\% & 1.7\% \\
        Corrections to simulations & 1.3\% & 4.3\% & 4.3\% \\
        \hline
        Total & 3.8\% & 5.4\% & 6.5\%\\
        \hline
    \end{tabular}
    \label{tab:uncertain}
\end{table}


\subsection{Uncertainties on the $H_b$ mass measurements}

The uncertainties on the mass and mass difference measurements come from the invariant-mass fit model, the momentum scale calibration, and the uncertainties on the $\Xic$ and $\Dsm$ masses~\cite{PDG2022}.
They are summarised in Table~\ref{tab:uncertain_mass} and Table~\ref{tab:uncertain_deltamass}.

The $H_b$ mass determined from the fit to the invariant-mass distribution is affected by the imperfect modelling of the signal, the combinatorial background and the partially reconstructed background.
Variations of the model for each fit component are studied in the same way as for the determination of the uncertainties on the signal yield described in Sec.~\ref{sec:uncertain_mass_fit}.
The largest variation of the mass obtained in these alternative fits compared to the nominal one is considered as the systematic uncertainty, which is $0.02$, $0.19$ and $0.09\mevcc$ for $m_{\Lb}$, $m_{\Xibz}$ and $m_{\Xibm}$, respectively. The larger uncertainty for $m_{\Xibz}$ is due to the higher background level.

Due to effects such as an imperfect alignment of the tracking system and the uncertainty on the magnetic field, the measured track momenta need to be calibrated to correct for possible biases. The calibration is performed using the masses of known hadrons~\cite{LHCb-PAPER-2011-035,LHCb-PAPER-2013-011} with a precision of $0.03\%$. The uncertainty is propagated to the $H_b$ mass measurement by varying the calibration by $\pm1$ standard deviation. Half of the difference between the two corresponding new $H_b$ masses is taken as the systematic uncertainty. The result, about $0.4\mevcc$, approximately scales with the energy release of the decay as $(m(H_b)-m(H_c)-m(\Dsm))\times0.03\%$. The uncertainty due to momentum scale calibration is assumed to be fully correlated for the three $H_b$ masses.

As mentioned in Sec.~\ref{sec:fit}, the $H_b$ invariant mass is calculated with the \Dsm and $H_c$ masses constrained to their previous world averages~\cite{PDG2022}.
The systematic uncertainty due to the $H_c$ and \Dsm masses is 0.16, 0.24, and 0.29 $\mevcc$ for the $\Lb$, $\Xibz$, and $\Xibm$ mass measurement, respectively. 
When measuring the mass difference between two different $H_b$ states, the uncertainty on the $\Dsm$ mass is cancelled. The remaining uncertainty on the $H_c$ mass varies between 0.23 and 0.31 $\mevcc$ depending on mass difference.

\begin{table}[tb]
    \centering
    \caption{Systematic uncertainties for the $H_b$ mass measurements.}
    \begin{tabular}{l r@{.}l r@{.}l r@{.}l}
        \hline
        Source & \multicolumn{2}{c}{\makecell[c]{$m_{\Lb}$\\$[\mevcc]$}} & \multicolumn{2}{c}{\makecell[c]{$m_{\Xibz}$\\$[\mevcc]$}} & \multicolumn{2}{c}{\makecell[c]{$m_{\Xibm}$\\$[\mevcc]$}}\\
        \hline
        Mass fit model & $\quad$0&02 & $\quad$0&19 & $\quad$0&09 \\
        Momentum scale calibration & 0&44 & 0&41 & 0&48 \\
        Uncertainties on the $H_c$ and $\Dsm$ masses & 0&16 & 0&24 & 0&29 \\
        \hline
    \end{tabular}
    \label{tab:uncertain_mass}
\end{table}

\begin{table}[tb]
    \centering
    \caption{Systematic uncertainties for the $H_b$ mass-difference measurements.}
    \begin{tabular}{l r@{.}l r@{.}l r@{.}l}
        \hline
        Source & \multicolumn{2}{l}{\makecell[c]{$m_{\Xibz}-m_{\Lb}$\\$[\mevcc]$}} & \multicolumn{2}{c}{\makecell[c]{$m_{\Xibm}-m_{\Lb}$\\$[\mevcc]$}}& \multicolumn{2}{c}{\makecell[c]{$m_{\Xibm}-m_{\Xibz}$\\$[\mevcc]$}}\\
        \hline
        Mass fit model & $\quad\ \ $0&19 & $\quad\ \ $0&09 & $\quad\ \ $0&21\\
        Momentum scale calibration & 0&03 & 0&04 & 0&07 \\
        Uncertainties on the $H_c$ mass & 0&27& 0&31& 0&23\\
        \hline
    \end{tabular}
    \label{tab:uncertain_deltamass}
\end{table}


\section{Results}%
\label{sec:results}

\begin{figure}[tb]
    \centering
    \includegraphics[width = 0.98\textwidth]{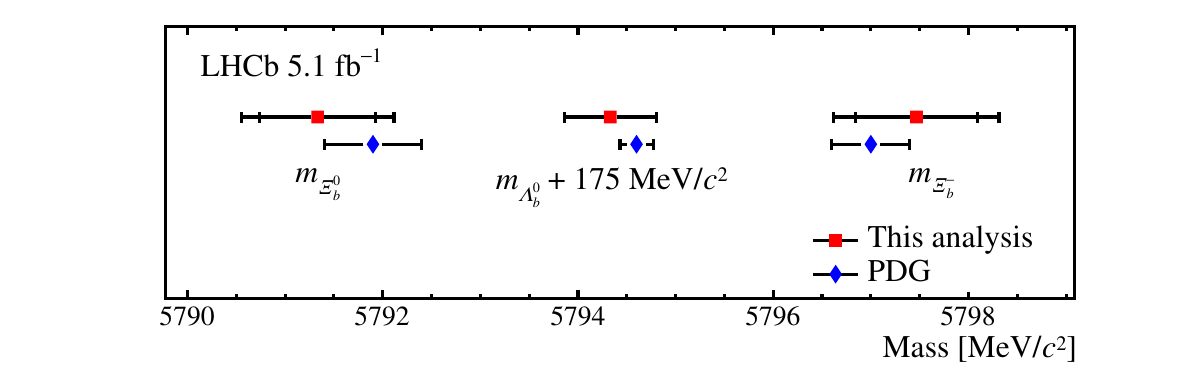}
    \caption{Comparison of measured (red) $b$ baryon masses with (blue) the PDG values~\cite{PDG2022}. The mass of \Lb is shifted upward by 175\mevcc to reduce the range of this plot. 
    The inner (outer) error bar is for the statistical (total) uncertainty.}
    \label{fig:m}
\end{figure}

Using the results presented in the previous sections, 
the $H_b$ masses and mass differences are measured to be
  \begin{align*}
      m_{\Lb}&=5619.34\pm0.06\pm0.44\pm0.16\mevcc,\\
      m_{\Xibz}&=5791.12\pm0.60\pm0.45\pm0.24\mevcc,\\
      m_{\Xibm}&=5797.02\pm0.63\pm0.49\pm0.29\mevcc,\\
      m_{\Xibz}-m_{\Lb}&=171.78\pm0.60\pm0.19\pm0.27\mevcc,\\
      m_{\Xibm}-m_{\Lb}&=177.68\pm0.63\pm0.10\pm0.31\mevcc,\\
      m_{\Xibm}-m_{\Xibz}&=5.90\pm0.87\pm0.22\pm0.23\mevcc,
  \end{align*}
where the first uncertainties are statistical, the second systematic, 
and the third due to those on masses of \Lc, \Xicp, \Xicz, and \Dsm hadrons. The measurements are consistent with previous world averages~\cite{PDG2022}, and comparisons are shown in Table~\ref{tab:result_vs_PDG} and Fig~\ref{fig:m}.

\begin{table}[tb]
    \centering
    \caption{Measured $H_b$ masses and mass differences and the previous world averages~\cite{PDG2022}.}
    \begin{tabular}{l c c}
        \hline \\ [-2.3ex]
         & This analysis [\mevcc] & Previous world average [\mevcc]\\ [0.3ex]
        \hline 
        $m_{\Lb}$   & $5619.34\pm0.47$   & $5619.60\pm0.17$\\
        $m_{\Xibz}$ & $5791.1\pm0.8$   & $5791.9\pm0.5$\\
        $m_{\Xibm}$ & $5797.0\pm0.8$   & $5797.0\pm0.6$\\
        $m_{\Xibz}-m_{\Lb}$   & $\,\,\,171.8\pm0.7$ & $\,\,\,172.5\pm0.4$\\
        $m_{\Xibm}-m_{\Lb}$   & $\,\,\,177.7\pm0.7$ & $\,\,\,177.46\pm0.31$\\
        $m_{\Xibm}-m_{\Xibz}$ & $\quad\,\,\: 5.9\pm0.9$ & $\quad\,\,\:5.9\pm0.6$\\
        \hline
    \end{tabular}
    \label{tab:result_vs_PDG}
\end{table}

\begin{figure}[tb]
    \centering
    \includegraphics[width = 0.8\textwidth]{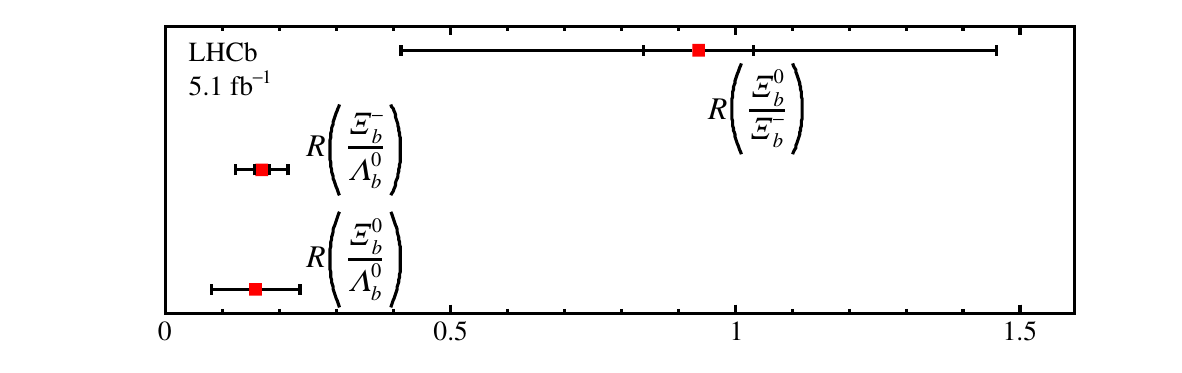}
    \caption{Measured $\mathcal{R}$ values. The inner (outer) error bar is for the statistical (total) uncertainty.}
    \label{fig:BF}
\end{figure}

The relative production rates of the three $H_b \to H_c D_s$ decays, 
given in Eq.~\ref{eq:R_xibz_lb}-~\ref{eq:R_xibz_xibm}, are measured to be 
\begin{align*}
  \mathcal{R}\left(\frac{\Xibz}{\Lb}\right)&=
  (15.8\pm1.1\pm0.6\pm7.7)\%,\\
  \mathcal{R}\left(\frac{\Xibm}{\Lb}\right)&=
  (16.9\pm1.3\pm0.9\pm4.3)\%,\\
  \mathcal{R}\left(\frac{\Xibz}{\Xibm}\right)&=
  (93.6\pm9.6\pm6.1\pm51.0)\%,
\end{align*}
where the first uncertainties are statistical, the second systematic, 
and the third due to those on the branching fractions of
$\Lc$, $\Xicp$, and $\Xicz$ decays. 
Figure~\ref{fig:BF} shows the measured $\mathcal{R}$ values.
The results are consistent with the $SU(3)$ flavour symmetry and predictions of phenomenological models~\cite{Hsiao:2015txa,Jiang:2018iqa}.

\section{Summary}
In this analysis, the dicharm decays of $\Xires_b$ baryons $\Xibz \to \Xicp \Dsm$ and $\Xibm \to \Xicz \Dsm$ are observed for the first time, using proton-proton collision data collected by the LHCb experiment at a centre-of-mass energy of $\sqrt{s} = 13\tev$, corresponding to an integrated luminosity of  5.1\invfb.
The masses of the \Lb, \Xibz and \Xibm baryons are measured through these two decays, and are consistent with known values~\cite{PDG2022}. These measurements will improve the world averages.
The relative branching fractions of these two decays are also measured.
The results are consistent with $SU(3)$ flavour symmetry and several predictions for relative production rates and decay branching fractions of $b$ baryons~\cite{Chua:2019yqh,Hsiao:2015txa,Li:2021kfb,Jiang:2018iqa}.
\clearpage

\section*{Acknowledgements}
%
%
\noindent We express our gratitude to our colleagues in the CERN
accelerator departments for the excellent performance of the LHC. We
thank the technical and administrative staff at the LHCb
institutes.
We acknowledge support from CERN and from the national agencies:
CAPES, CNPq, FAPERJ and FINEP (Brazil); 
MOST and NSFC (China); 
CNRS/IN2P3 (France); 
BMBF, DFG and MPG (Germany); 
INFN (Italy); 
NWO (Netherlands); 
MNiSW and NCN (Poland); 
MEN/IFA (Romania); 
MICINN (Spain); 
SNSF and SER (Switzerland); 
NASU (Ukraine); 
STFC (United Kingdom); 
DOE NP and NSF (USA).
We acknowledge the computing resources that are provided by CERN, IN2P3
(France), KIT and DESY (Germany), INFN (Italy), SURF (Netherlands),
PIC (Spain), GridPP (United Kingdom), 
CSCS (Switzerland), IFIN-HH (Romania), CBPF (Brazil),
Polish WLCG  (Poland) and NERSC (USA).
We are indebted to the communities behind the multiple open-source
software packages on which we depend.
Individual groups or members have received support from
ARC and ARDC (Australia);
Minciencias (Colombia);
AvH Foundation (Germany);
EPLANET, Marie Sk\l{}odowska-Curie Actions, ERC and NextGenerationEU (European Union);
A*MIDEX, ANR, IPhU and Labex P2IO, and R\'{e}gion Auvergne-Rh\^{o}ne-Alpes (France);
Key Research Program of Frontier Sciences of CAS, CAS PIFI, CAS CCEPP, 
Fundamental Research Funds for the Central Universities, 
and Sci. \& Tech. Program of Guangzhou (China);
GVA, XuntaGal, GENCAT, Inditex, InTalent and Prog.~Atracci\'on Talento, CM (Spain);
SRC (Sweden);
the Leverhulme Trust, the Royal Society
 and UKRI (United Kingdom).


\section*{Appendices}

\appendix

\section{Non-dicharm contribution}
\label{sec:non-dicharm}
\begin{figure}[b]
    \centering
    \includegraphics[width=0.49\linewidth]{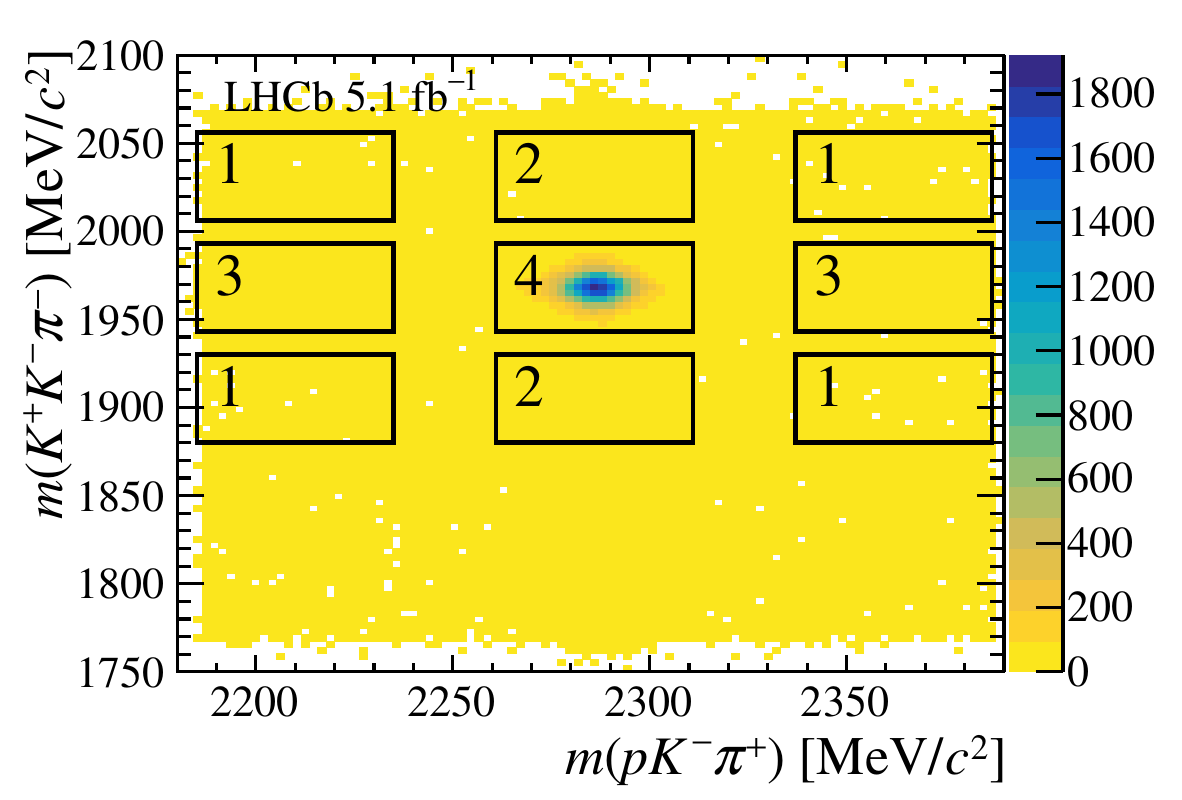}
    \includegraphics[width=0.49\linewidth]{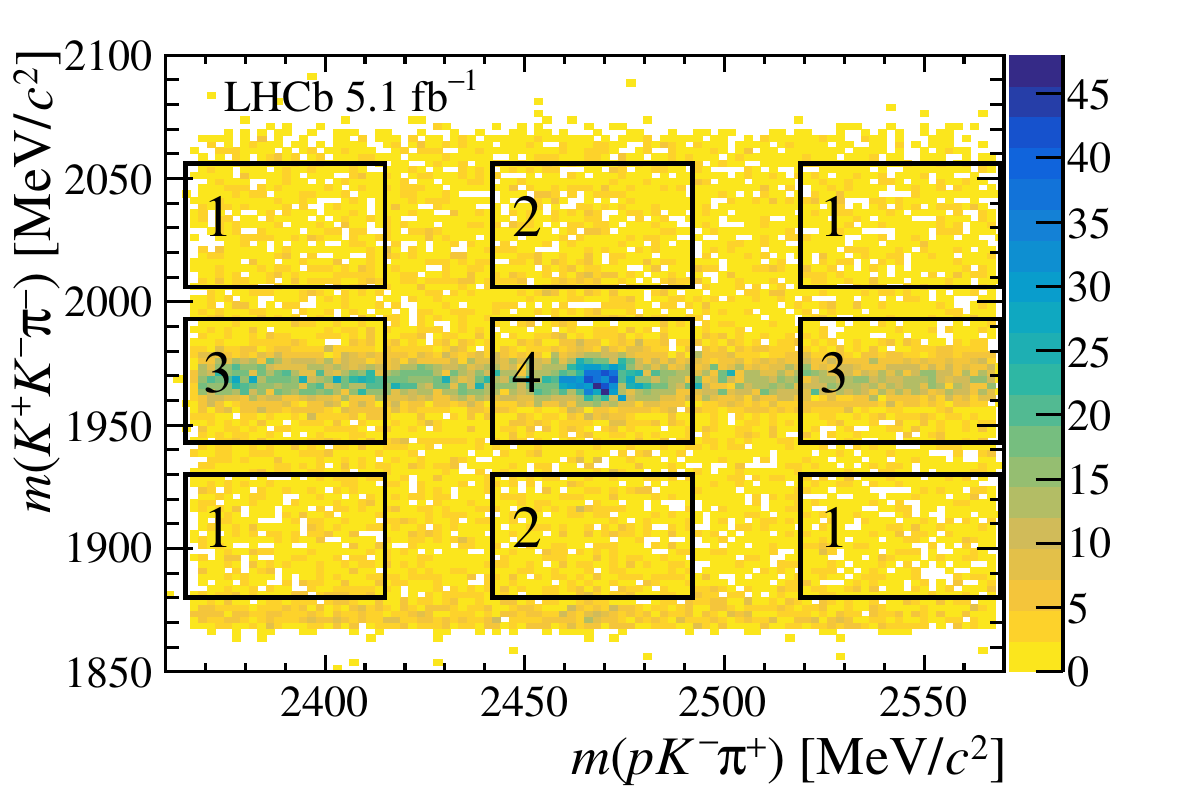}
    \includegraphics[width=0.49\linewidth]{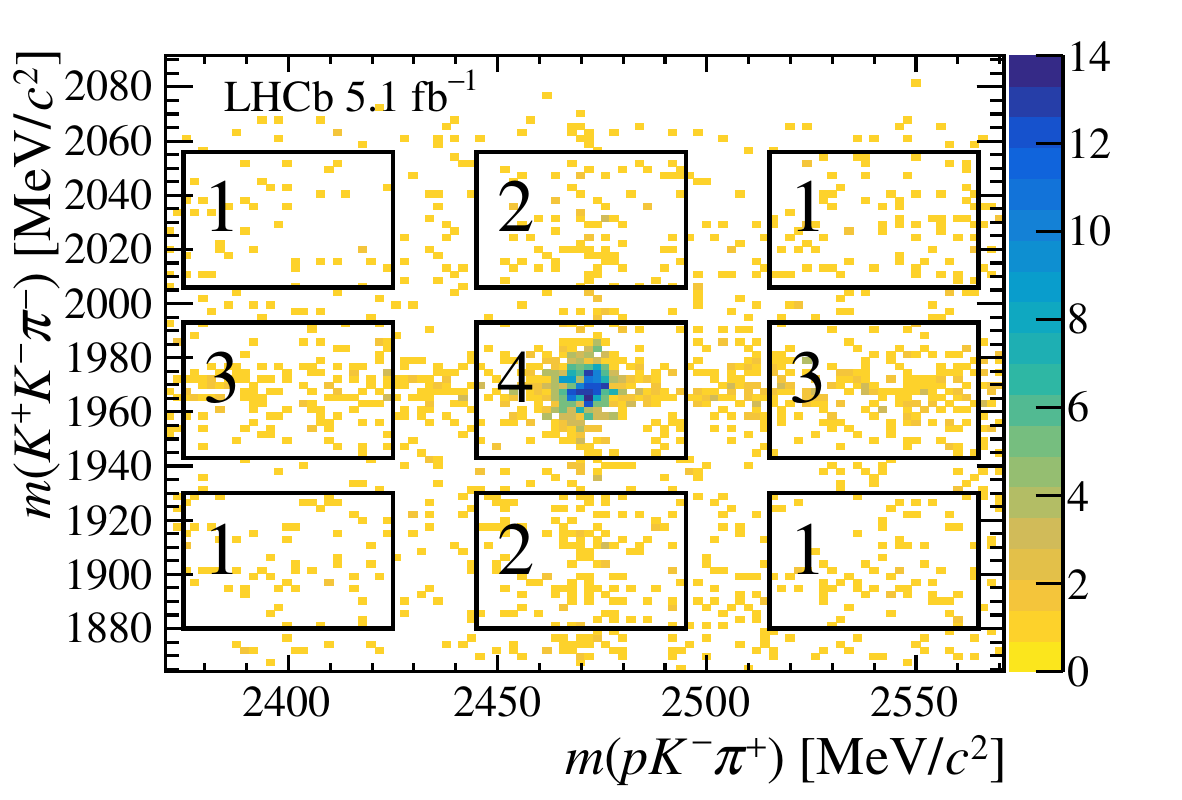}
    \caption{Distributions of the $H_c$ mass versus the $\Dsm$ mass with the regions 1--4 indicated. The regions illustrate the signal region and the $H_c$ and/or \Dsm sideband regions.}
    \label{fig:mDs_mXc}
\end{figure}

\begin{figure}[tb]
    \centering
    \includegraphics[width=0.48\linewidth]{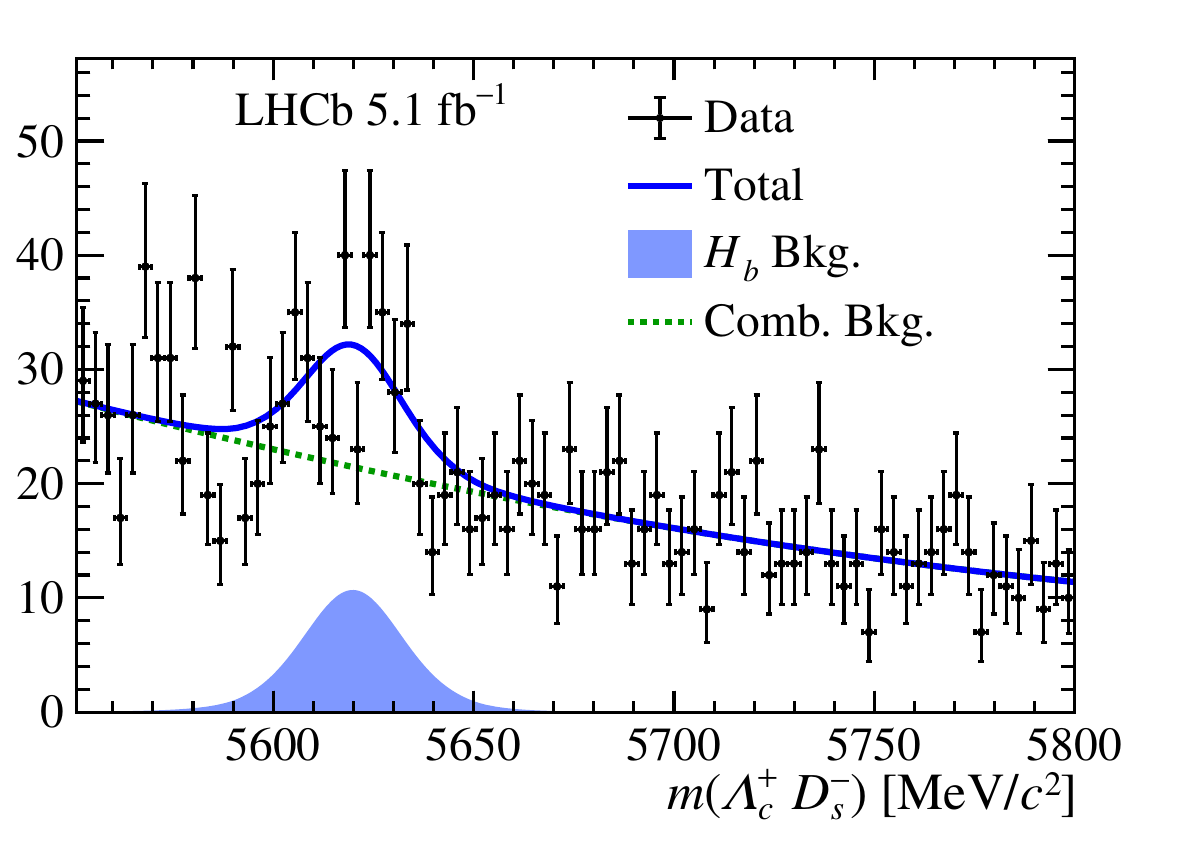}
    \includegraphics[width=0.48\linewidth]{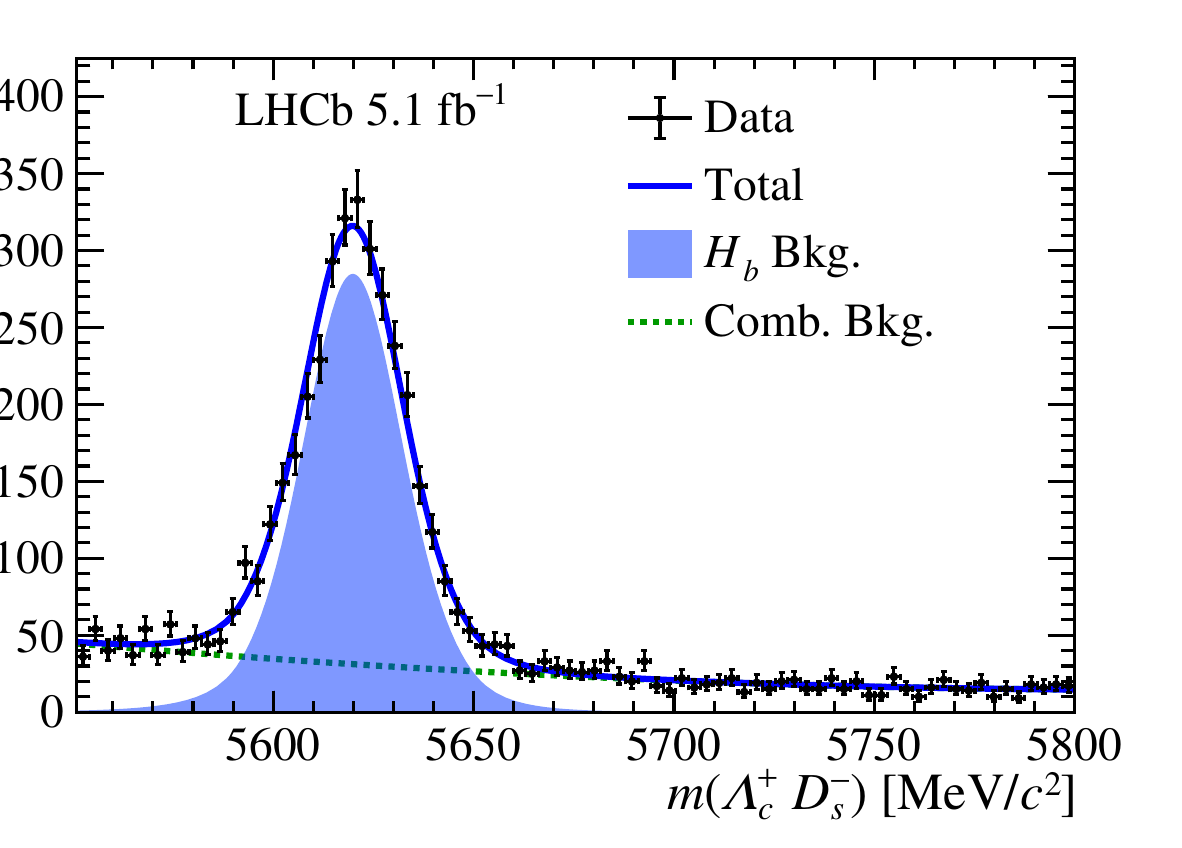}\\
    \includegraphics[width=0.48\linewidth]{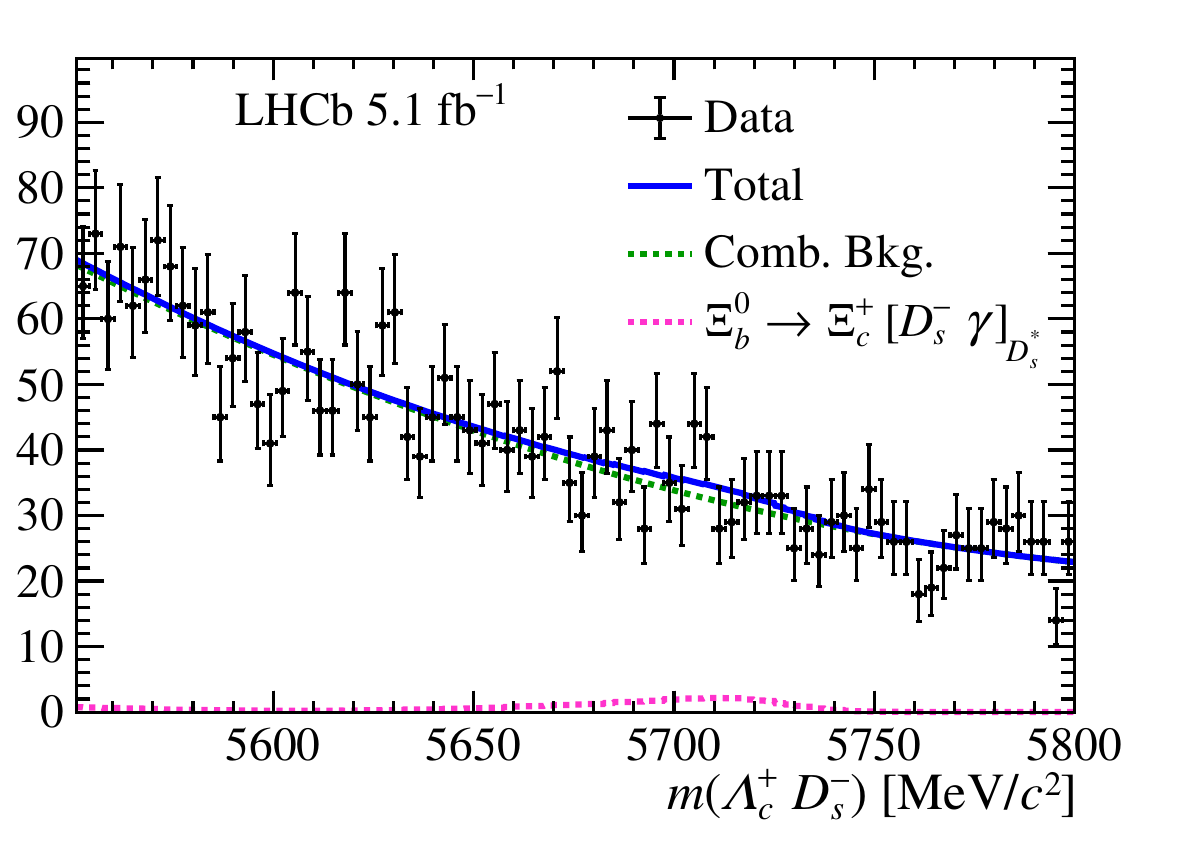}
    \caption{Invariant-mass distributions of \Lb candidates in the 
    (top left) region 1, 
    (top right) region 2, 
    and (bottom) region 3.}
    \label{fig:fit_charmless_lb}
\end{figure}

\begin{figure}[tb]
    \centering
    \includegraphics[width=0.48\linewidth]{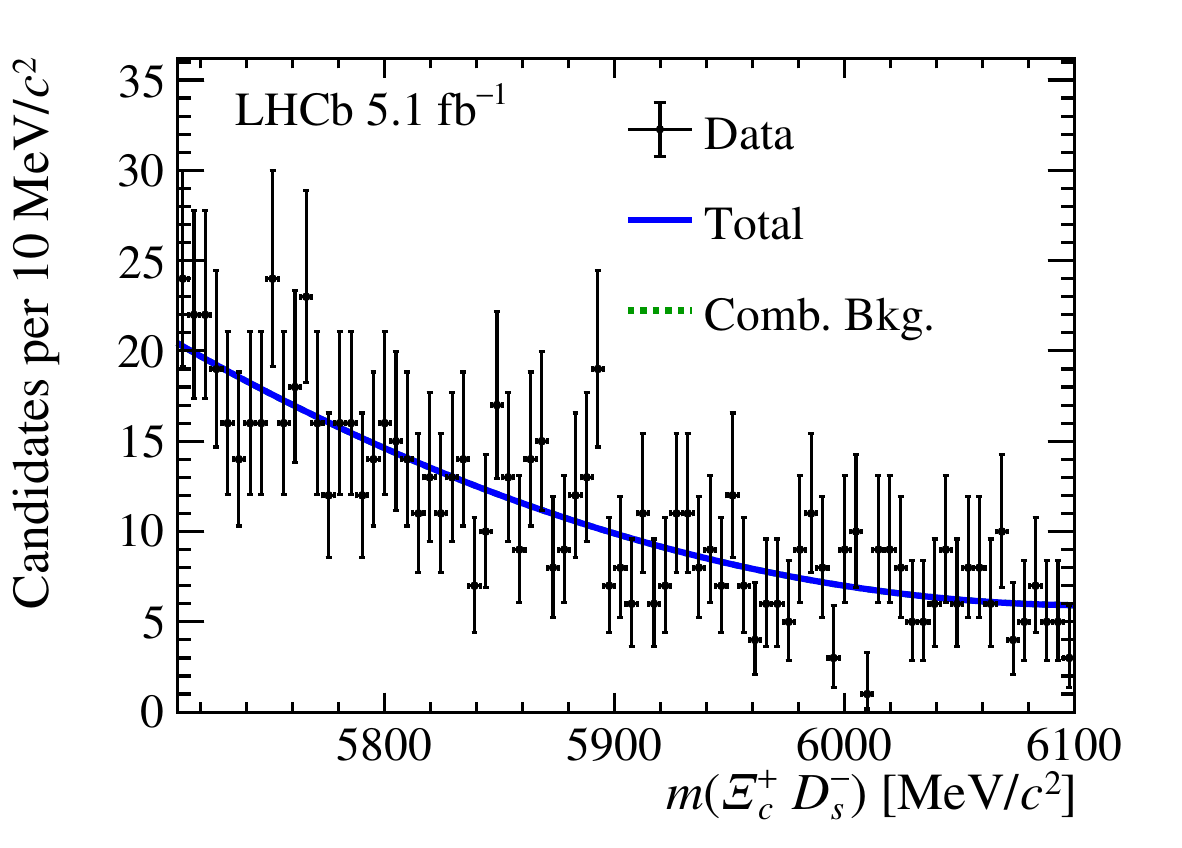}
    \includegraphics[width=0.48\linewidth]{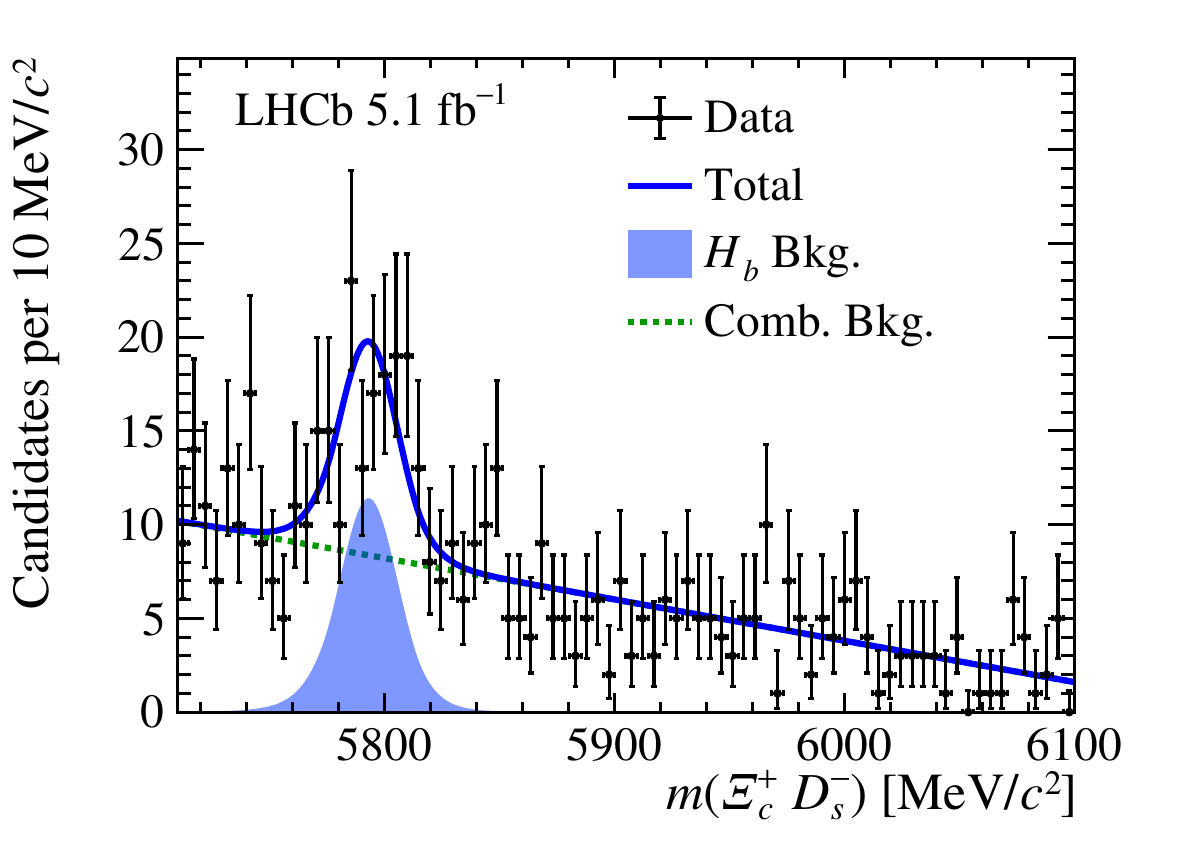}\\
    \includegraphics[width=0.48\linewidth]{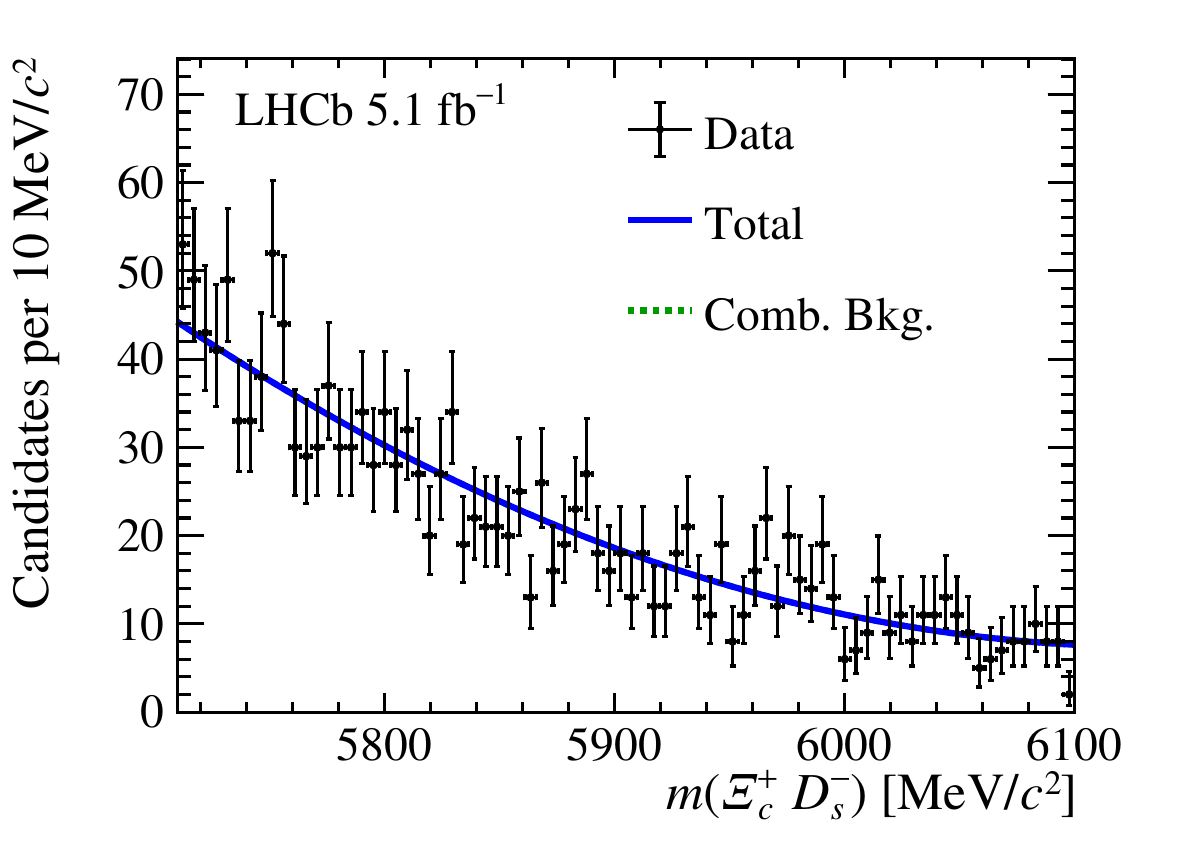}
    \caption{Invariant-mass distributions of \Lb candidates in the 
    (top left) region 1, 
    (top right) region 2, 
    and (bottom) region 3.}
    \label{fig:fit_charmless_xib0}
\end{figure}

\begin{figure}[tb]
    \centering
    \includegraphics[width=0.48\linewidth]{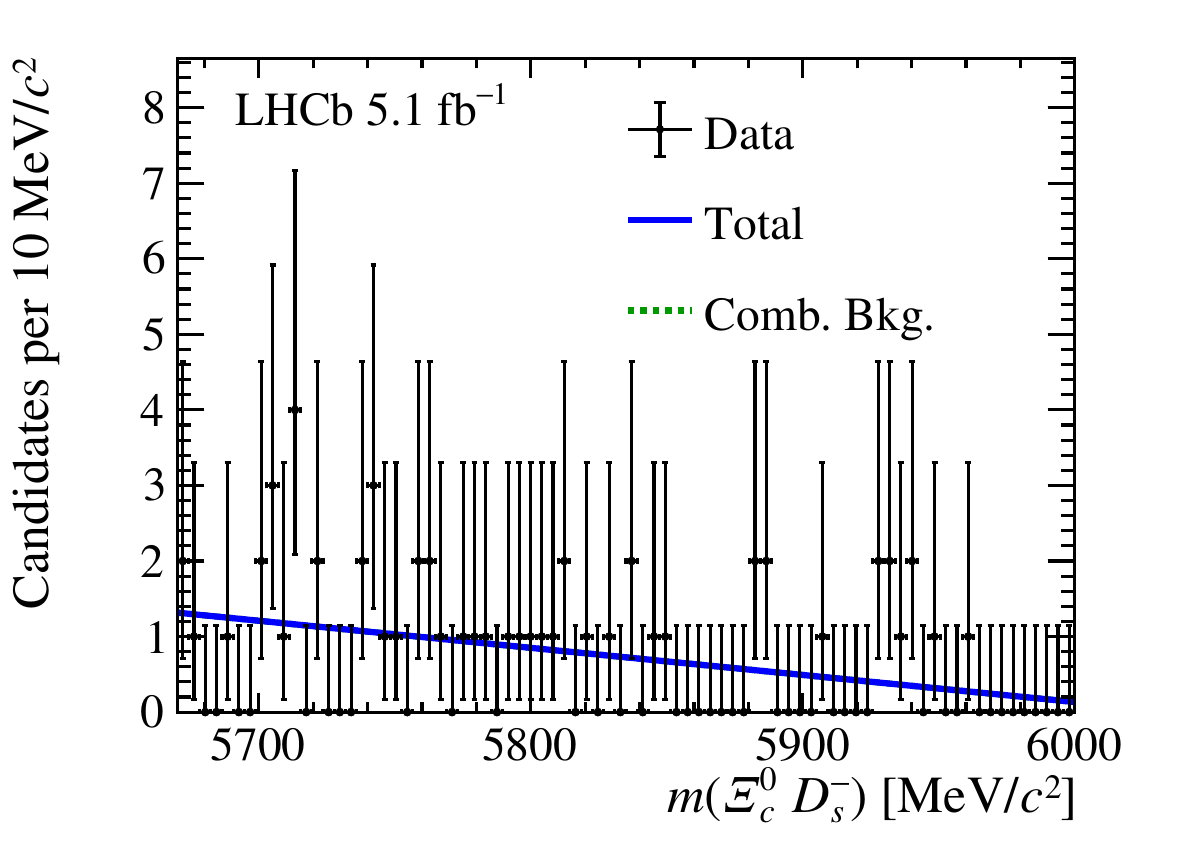}
    \includegraphics[width=0.48\linewidth]{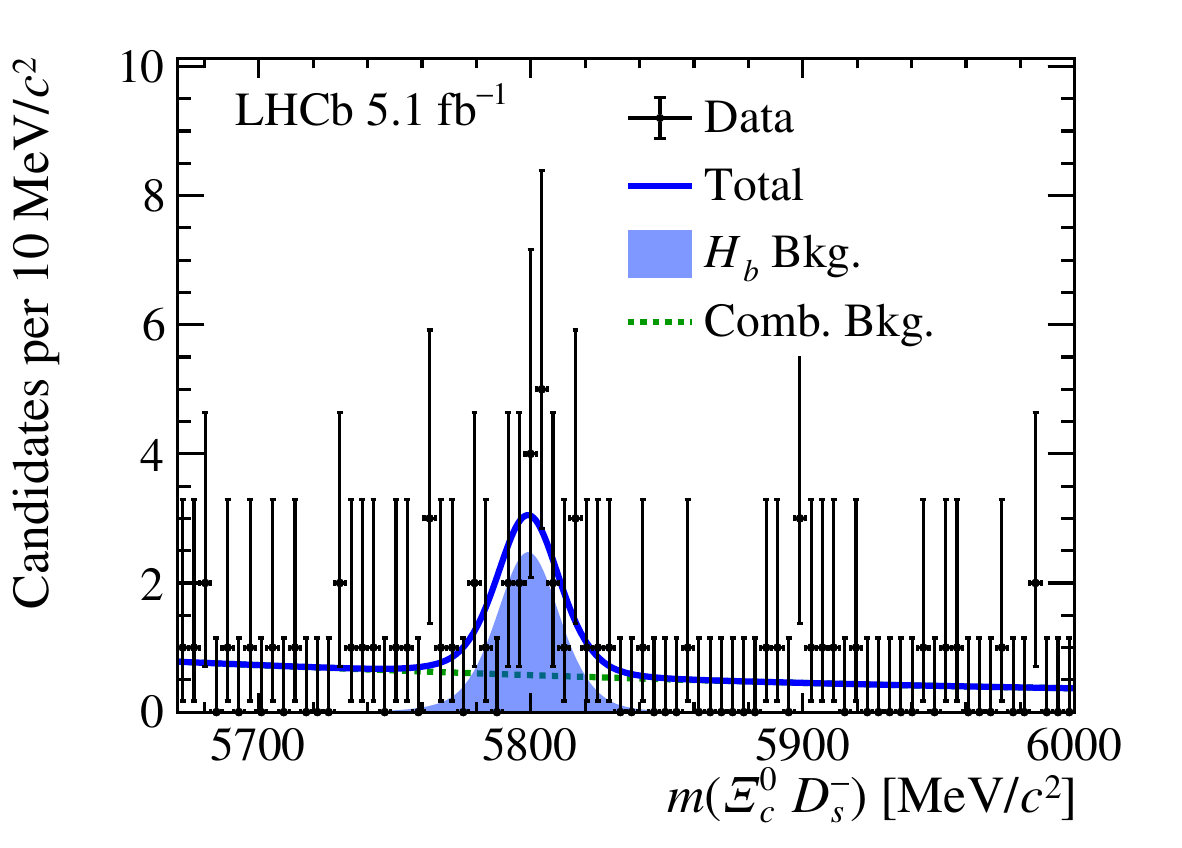}\\
    \includegraphics[width=0.48\linewidth]{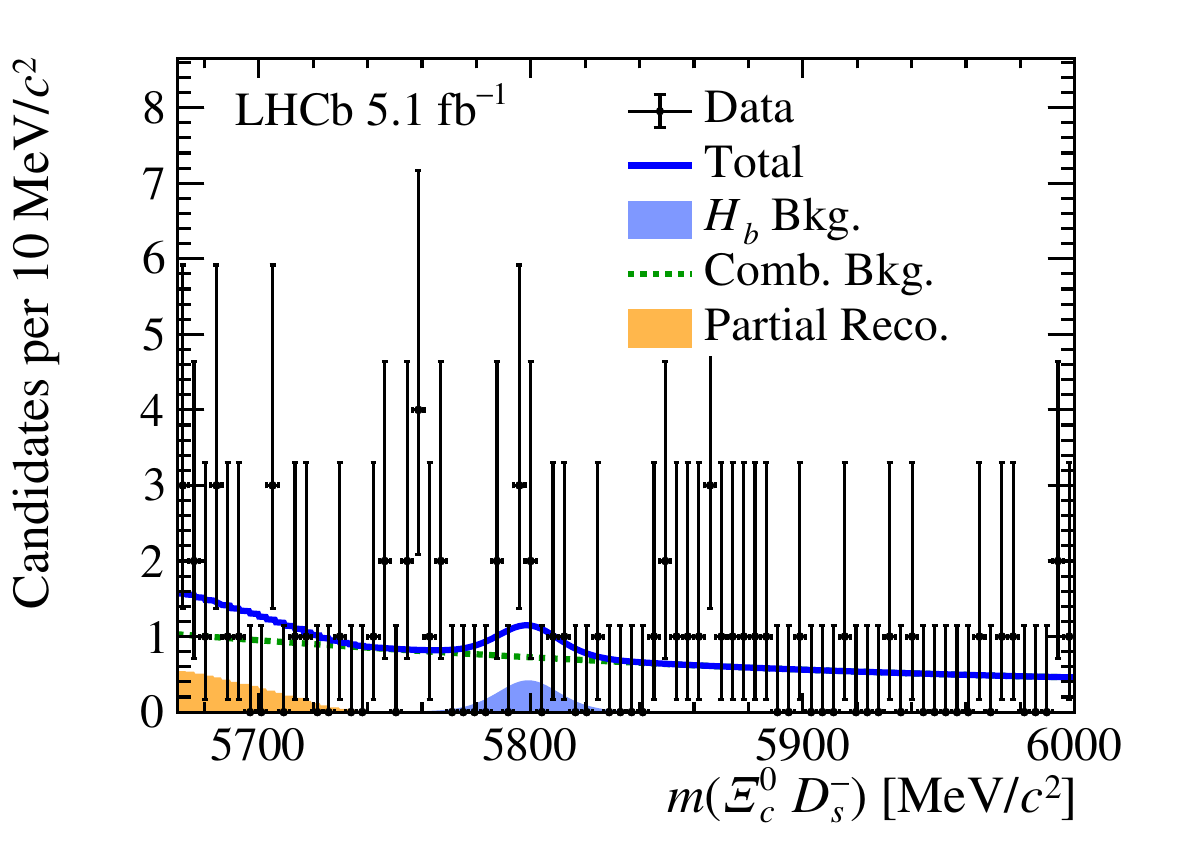}
    \caption{Invariant-mass distributions of \Lb candidates in the 
    (top left) region 1, 
    (top right) region 2, 
    and (bottom) region 3.}
    \label{fig:fit_charmless_xibm}
\end{figure}

Three distinct sources of non-dicharm backgrounds are considered: 
\begin{itemize}
    \item The $H_b\to (p\Km(\Km)\pip)(\Kp\Km\pip)$ decay with neither the $H_c$ nor the $\Dsm$ hadrons.
    \item The $H_b\to (p\Km(\Km)\pip)\Dsm$ decay without the $H_c$ baryon.
    \item The $H_b\to H_c(\Kp\Km\pip)$ decay without the $\Dsm$ meson.
\end{itemize}
Figure~\ref{fig:mDs_mXc} shows the two-dimensional $H_c$ versus \Dsm invariant-mass distribution in the signal region and the $H_c$ and/or \Dsm sideband regions. There are four regions illustrated in Fig~\ref{fig:mDs_mXc}: 
\begin{itemize}
    \item The region 1 lies in the $H_c$ and \Dsm sideband region.
    \item The region 2 lies in the $H_c$ signal and \Dsm sideband region. 
    \item The region 3 lies in the $H_c$ sideband and \Dsm signal region. 
    \item The region 4 lies in the $H_c$ and \Dsm signal region.
\end{itemize}
The $H_b$ signal yields in the $H_c$ and/or \Dsm sideband regions are estimated by simultaneous fitting to the $H_b$ invariant-mass spectra in these four regions. 
The fit model is similar as the one mentioned in Sec.~\ref{sec:fit}. 
Figures~\ref{fig:fit_charmless_lb},~\ref{fig:fit_charmless_xib0}, and~\ref{fig:fit_charmless_xibm} show the \Lb, \Xibz, and \Xibm invariant-mass distributions in the $H_c$ and/or \Dsm sideband regions superimposed by the fit results, respectively. 

\clearpage

\addcontentsline{toc}{section}{References}
\bibliographystyle{LHCb}
\bibliography{main,LHCb-PAPER,LHCb-CONF,LHCb-DP, standard,Mybib}

\newpage
\newpage
\centerline
{\large\bf LHCb collaboration}
\begin
{flushleft}
\small
R.~Aaij$^{33}$\lhcborcid{0000-0003-0533-1952},
A.S.W.~Abdelmotteleb$^{52}$\lhcborcid{0000-0001-7905-0542},
C.~Abellan~Beteta$^{46}$,
F.~Abudin{\'e}n$^{52}$\lhcborcid{0000-0002-6737-3528},
T.~Ackernley$^{56}$\lhcborcid{0000-0002-5951-3498},
B.~Adeva$^{42}$\lhcborcid{0000-0001-9756-3712},
M.~Adinolfi$^{50}$\lhcborcid{0000-0002-1326-1264},
P.~Adlarson$^{78}$\lhcborcid{0000-0001-6280-3851},
H.~Afsharnia$^{10}$,
C.~Agapopoulou$^{44}$\lhcborcid{0000-0002-2368-0147}
% ...

{\footnotesize \it

% ...
$^{10}$Universit{\'e} Clermont Auvergne, CNRS/IN2P3, LPC, Clermont-Ferrand, France\\
% ...
$^{33}$Nikhef National Institute for Subatomic Physics, Amsterdam, Netherlands\\
% ...
$^{42}$Instituto Galego de F{\'\i}sica de Altas Enerx{\'\i}as (IGFAE), Universidade de Santiago de Compostela, Santiago de Compostela, Spain\\
% ...
$^{44}$European Organization for Nuclear Research (CERN), Geneva, Switzerland\\
% ...
$^{46}$Physik-Institut, Universit{\"a}t Z{\"u}rich, Z{\"u}rich, Switzerland\\
% ...
$^{50}$H.H. Wills Physics Laboratory, University of Bristol, Bristol, United Kingdom\\
% ...
$^{52}$Department of Physics, University of Warwick, Coventry, United Kingdom\\
% ...
$^{55}$School of Physics and Astronomy, University of Glasgow, Glasgow, United Kingdom\\
$^{56}$Oliver Lodge Laboratory, University of Liverpool, Liverpool, United Kingdom\\
% ...
$^{78}$Department of Physics and Astronomy, Uppsala University, Uppsala, Sweden, associated to $^{55}$\\
% ...
}
\end{flushleft}

%
%
%
%

\end{document}